**Multiscale Modeling of Zr Corrosion: Predicting Microstructures of Oxide Films**

**Authors**: K. Starkey, L. Kahle, M. Christensen, V. Eyert, A. El-Azab, C. B. Geller, A. P. Shivprasad, and E. Wimmer

**Abstract**

A first-principles-informed multiscale modeling approach is reported for oxide formation during corrosion of zirconium. The present simulations reproduce the columnar corrosion film microstructure and grain size distribution seen in micrographs of oxidized Zr surfaces. This structural feature emerges from a hexagonal ZrO layer adjacent to the $\alpha$-Zr/$ZrO_{2-x}$ interface, which is dynamically stabilized by continuous heterogeneous grain nucleation and grain boundary mass diffusion. This ZrO layer is identified as a significant oxygen diffusion barrier and a potential contributor to sub-parabolic pretransition corrosion kinetics. The simulations also reproduce the undulatory nature of the metal/oxide interface, leading to regions with tensile stress components capable of nucleating cracks, despite the overall state of compressive hydrostatic stress within the oxide film. First principles density functional theory (DFT) computations of fundamental $ZrO_x$ properties were combined with molecular dynamics simulations using a machine-learned potential based on a DFT training set to provide extensive input to an advanced phase field model for oxide formation and its microstructure, as implemented in the new *MedeA PhaseField* module. The phase field simulations reported herein incorporated computed stoichiometry-dependent thermochemical data, anisotropic elastic properties, interface energies, bulk and grain boundary mass transport processes, multiphase heterogeneous grain nucleation, and strain energy-dependent phase stabilities.

## 1. Introduction

Zirconium's low thermal neutron capture cross-section [1], favorable thermo-mechanical properties [2], and aqueous corrosion resistance [3,4] have made Zr-based alloys the materials of choice for pressurized light-water reactor (LWR) fuel claddings throughout the nuclear power industry [5]. Nevertheless, aqueous corrosion of Zr alloys and the associated hydrogen pickup (HPU) often limit the safe operating lifetimes and operating temperatures of LWR fuel elements [6]. Degradation of fuel cladding by oxidation during accident conditions could augment the risk of ballooning and bursting by further embrittling the cladding, resulting in the release of fission products to the coolant [5]. Furthermore, hydrogen picked up by the cladding as a result of aqueous corrosion forms zirconium hydrides, which can degrade fuel cladding mechanical integrity and, thus, fission product containment [7,8].

Uncertainties in Zr alloy aqueous corrosion and HPU behaviors as they relate to LWR service conditions and alloy chemistry continue to impose reactor design and operational penalties even after over half a century of empirical engineering-oriented testing [9]. Models for oxide formation during aqueous corrosion of Zr and Zr alloys were developed in conjunction with these tests, which focused on the prediction of oxide layer thickness and hydrogen transport through the oxide [10]. However, these empirical models do not explicitly account for variations in the complex, evolving microstructures of Zr alloy aqueous corrosion scales, which include extensive lateral cracking and periodic loss of protective character.

Several factors limit the accuracy of cladding corrosion and HPU predictions, including the necessary duration and expense of prototypical in-pile testing, the variability of expected in-pile service histories (including possible accidents), and the imprecise knowledge and intrinsic complexity of the multiscale chemo-mechanical material processes governing aqueous corrosion and HPU behavior [11]. Recent work emphasized that a fundamental analytical description of Zr aqueous corrosion with predictive capability remains elusive despite extensive empirical investigation, owing to numerous strongly coupled multiscale physical and chemical processes [12].

Previous phase field models of Zr oxidation treated the oxide layer as a homogeneous monolithic phase with directionally averaged elastic properties [12,16], and similar approaches have addressed multilayer oxide growth, dose rate effects, and incipient hydride formation [13,14,15]. While these models predict periodic oxide layer breakdown [16], they do not resolve the microstructure that governs corrosion kinetics and that can act as deep traps for hydrogen.

Zirconium and its alloys, such as the Zircaloys and Nb-containing alloys (e.g., Zr-2.5Nb, M5, E110, ZIRLO, and E635) exhibit several microstructural characteristics that a simulation tool should endeavor to reproduce. First, aqueous corrosion films that grow on Zr alloys consist predominantly of monoclinic $ZrO_2$ (m-$ZrO_2$) but contain regions of both tetragonal $ZrO_2$ (t- $ZrO_2$) and hexagonal ZrO (h-ZrO). The h-ZrO constituent of the oxide consists of clusters of small grains extending along and near the metal/oxide interface [17], indicating that h-ZrO is a transitional phase. The volume fraction of the t-$ZrO_2$ constituent of oxide films may vary both spatially and temporally in a periodic manner [18].

Second, the growing oxide films develop a predominant crystallographic texture with a primarily columnar microstructure, where the column axes are roughly parallel to the direction of oxidation, while the nascent grains near the metal/oxide interface tend to be equiaxed. The metal/oxide interface has an undulating nature often described as having a “cauliflower” texture, due to the fine scale of the oxide grains at the oxide-metal interface [19]. Interface undulations are shown to be capable of producing adjacent regions with tensile stress components, despite the overall compressive state of the film. It was also observed that the oxide films exhibit distributions of fine intergranular pores and cracks, which may coalesce into continuous vertical transport paths for both oxygen (O) and hydrogen (H). The potential role for HPU of such nano-scale features near the metal/oxide interface has been pointed out previously [20].

The current work focuses on reproducing the pretransition corrosion film microstructure only (hydrogen pickup is not modeled). Furthermore, effects of alloying additions, irradiation, and prototypical in-reactor water chemistries are deferred to later work. A multiscale modeling approach to Zr alloy oxidation is adopted, including state-of-the-art electronic structure computations, atomistic molecular dynamics simulations employing machine-learning technology, and a novel microstructural evolution modeling tool, called *MedeA PhaseField* [21], based on phase field methodology. The current work advances beyond the 1D interface tracking and effective-layer models described above, resolving grain-scale microstructures and anisotropic stress fields within a multiphysics, multiscale phase field framework. Novel enhancements include refined heterogeneous nucleation sampling at interfaces and a grain boundary diffusion formulation independent of the artificial diffuse interface widths necessarily employed in phase field models. Thermodynamic, elastic, and kinetic parameters are derived from first-principles calculations and a purpose-built machine-learned interatomic potential trained on first principles computations, with predictions validated against experimental microstructural data.

This paper is organized as follows. Section 2 describes the computational methods, including density functional theory calculations, machine-learned potentials, and atomistic microstructure models. This section also presents the phase field model formulation, incorporating the grand potential equations, the implementation of stress coupling, the oxygen diffusion treatment, and the grain nucleation mechanisms. Section 3 reports calculated material properties and the phase field simulation results, including structural and thermodynamic properties, elastic properties, grain boundary and interphase energies, oxygen diffusion flux plots, oxide microstructural evolution, a nucleation analysis, and stress distributions of the oxide phases. This section also provides statistical analyses of computed microstructures and their experimental validation. Conclusions are then presented in Section 4.

## 2. Computational Approaches

Computational methods for calculating materials properties are described first, followed by phase field model formulation.

## Density Functional Theory

We use Kohn-Sham density functional theory (DFT) [22] to calculate electronic and structural properties from first principles. We employ the efficient plane-wave implementation of DFT of the Vienna Ab initio Simulation Package (VASP) [23,24] within the *MedeA* materials modeling environment [25], and solve the equations using the projector-augmented-wave (PAW) method [26]. The Perdew, Burke, Ernzerhof (PBE) exchange-correlation functional [27,28] with a plane-wave cutoff energy of 500 eV is used on a k-point mesh with a spacing of 0.2 $Å^{-1}$ with Gaussian smearing applied with a width of 0.05 eV. In elastic property computations, the plane-wave cutoff energy is increased to 520 eV and the k-point mesh density is set to 0.15 $Å^{-1}$. Equilibrium structures are determined using a conjugate gradient minimization technique with a convergence criterion of 0.01 eV/Å for residual forces.

## Machine-learned potentials

To access longer length and time scales, a machine-learned potential (MLP) is trained on DFT results to describe interatomic interactions in the Zr–O system. The Spectral Neighbor Analysis Potential (SNAP) with quadratic terms is employed using the FitSNAP code [29–31] within the *MedeA* MLPGenerator framework. The FitSNAP band limit parameter is set to 8, with optimized parameters yielding a radial cutoff of 4.747 Å, relative radii of 0.500 and 0.587 for Zr and O, and weights of 0.550 and 0.186 for Zr and O, respectively. The MLP is generated by fitting 631,684 data points (energies, forces, and stress tensors) from DFT computations on 4,006 structures. Energy, force, and stress components are weighted by 1.0, 0.05, and $10^{-5}$, respectively, yielding RMSE values of 0.018 eV/atom for energies, 0.014 eV/Å for forces, and 0.026 bar for stress.

The training set includes polymorphs of relevant phases. For elemental Zr, $\alpha$-Zr (hcp), $\omega$-Zr (hexagonal), $\beta$-Zr (bcc), and fcc-Zr are included. The training set also comprises O-dissolved in metallic phases in the dilute limit and saturated $\alpha$-Zr(O-sat) containing 25 at.% O, i.e., hexagonal $Zr_3O$ (hexagonal). Oxides included in the training set are the monoxides of ZrO (both hexagonal and fcc) and the dioxides of $ZrO_2$ in the cubic, tetragonal, monoclinic, and orthorhombic polymorphs. In addition to the equilibrium configurations of these structures, strained structures are also added to the training set. Both isotropic and uniaxial strains up to ±6% and angular deformations up to ± 2° are applied.

Furthermore, point defects, such as O-vacancies, and planar defects, including surfaces and interfaces, are included in the training set. Surface structures of the four polymorphs of elemental Zr and structures of $\alpha$-Zr with stacking faults were constructed and added to the training set as well. In addition to the single-phase structures, the training set also includes interface structures. Interfaces between α-Zr(O-sat) and all the different oxide structures are included, as well as h-ZrO/m-$ZrO_2$, h-ZrO/t-$ZrO_2$, t-$ZrO_2$/m-$ZrO_2$, and m-$ZrO_2$/m-$ZrO_2$ interfaces. For both single-phase structures and interface structures, training set structures are generated by high-temperature molecular dynamics simulations using the on-the-fly machine-learning capabilities of *MedeA VASP* [32]. This strategy is an efficient way to create training sets with minimal redundancy. A total of 3,604 Zr-O structures has been used to fit the MLP used subsequently in this work.

## Kinetic Monte Carlo

In the present work, the diffusivity of oxygen atoms in the relevant phases is obtained from kinetic Monte Carlo simulations. To this end, possible sites of the diffusing species are found from a geometric analysis. For each possible site, all nearest-neighbor sites are identified and the energy barriers, $E_b$, between these sites are computed using the nudged elastic band method with energies and forces obtained from DFT calculations. Jump rates are then computed using the expression $\Gamma = kT/h \times \exp([-E_b/(RT)]$ with $k$ being the Boltzmann constant, $T$, the absolute temperature in Kelvin, $h$ *is* Planck's constant, and $R$ us the universal gas constant. Using the connectivity between adjacent local minima and the corresponding jump rates, kinetic Monte Carlo simulations are then performed for different temperatures between 300 K and 1000 K, typically employing several hundred million Monte Carlo moves. The diffusivities are fitted to the Arrhenius form $D = D_0 \times \exp([-Q/(RT)]$ resulting in an effective activation barrier, $Q$, and a temperature-independent prefactor, $D_0$.

## Atomistic microstructure models

Interface energies between randomly oriented oxide grains are obtained through atomic microstructure models constructed using the *MedeA* Microstructure Builder, which employs the Voronoi tessellation to define grains from random seed points. Crystal seeds are positioned at these points and grow outward until reaching grain boundaries defined by Voronoi facets, with constraints imposed on interatomic distances to prevent atoms from approaching closer than the sum of atomic radii (0.547 Å for O and 1.086 Å for Zr). Grain orientations are sampled randomly and uniformly from the space of 3D rotations. The interface energy is calculated from the total energy of the grain boundary system, the total energy of the perfect crystal, and the interface area.

## Phase field

Phase-field methods iapproximate atomically sharp interfaces between adjacent phases or grains as diffuse regions of finite width, in which the order parameter defining the phase or grain varies smoothly and continuously from one to the other. [33,34]. As a continuum framework, the method effectively coarse-grains atomic-scale information, eliminating explicit interface tracking while naturally handling complex topological changes in evolving microstructures with multiple interfaces [35].

Grand Potential Equations

In the current work, we use a grand potential formulation of the thermodynamics leading to the phase field equations. The grand potential functional, $\Omega$, , of the system is expressed in the form:

$$\Omega = \int [\sum_{\alpha} \mathrm{h}^{\alpha}\omega^{\alpha} + \mathrm{f}_{\mathrm{int}} + \mathrm{f}_{\mathrm{elastic}}]\, \mathrm{dV}. \tag{1}$$

In the above expression, $\omega^{\alpha} = \mathrm{f}_{\mathrm{chem}}^{\alpha} - \mu\rho$ is the Legendre transformation of the chemical Helmholtz free energy, $\mathrm{f}_{\mathrm{chem}}^{\alpha}$, (the compositional dependent terms) with $\mu$ and $\rho$ being the oxygen chemical potential and number density of oxygen atoms, respectively, $\mathrm{f}_{\mathrm{int}}$ is the interfacial energy density, $\mathrm{f}_{\mathrm{elastic}}$ is the elastic energy density, and $\mathrm{h}^{\alpha}$ is an interpolation function given by

$$\mathrm{h}^{\alpha} = \frac{\sum_{\alpha} \eta_{\alpha i}^{2}}{\sum_{i} \eta_{i}^{2}} \tag{2}$$

where $\eta_{\alpha i}$is the order parameter representing grain $i$ in phase $\alpha$. The interfacial energy density $\mathrm{f}_{\mathrm{int}}$, is calculated as:

$$\mathrm{f}_{\mathrm{int}} = \frac{1}{2}\sum_{\alpha i} \kappa_{\eta_{\alpha i}} |\nabla\eta_{\alpha i}|^{2} + \mathrm{m}\left(\sum_{\alpha i} \frac{\eta_{\alpha i}^{4}}{4} - \frac{\eta_{\alpha i}^{2}}{2} + \eta_{\alpha i}^{2}\left(\sum_{\beta j \neq \alpha i} \gamma_{ij}(\nabla\boldsymbol{\eta})\eta_{\beta j}^{2}\right) + \frac{1}{4}\right). \tag{3}$$

The gradient parameters $\kappa_{\eta_{\alpha i}}$, and the model parameters $\mathrm{m}$ and $\gamma_{ij}$ can be computed straightforwardly from the surface energies and the diffuse interface width [36].

The chemical free energy density is interpolated amongst the various phases:

$$\mathrm{f}_{\mathrm{chemical}} = \sum_{\alpha} \mathrm{h}^{\alpha}\mathrm{f}_{\mathrm{chem}}^{\alpha} \tag{4}$$

where $\mathrm{f}_{\mathrm{chem}}^{\alpha}$ is the chemical free energy density for phase $\alpha$ and is written as a parabolic function of the O mole fraction, denoted by $c_{\mathrm{O}}$. We delay the definition of $\mathrm{f}_{\mathrm{elastic}}$, $\mathrm{f}_{\mathrm{fracture}}$, and $\mathrm{f}_{\mathrm{nucleation}}$ for the next sections.

The Allen-Cahn equations are [34]:

$$\frac{\partial \eta_{\alpha i}}{\partial t} = -L\frac{\delta\Omega}{\delta\eta_{\alpha i}} = -L\left[\frac{\partial}{\partial\eta_{\alpha i}}(\,\omega^{\alpha} + f_{int}) - \nabla\cdot\left(\kappa_{\eta_{\alpha i}}\nabla\eta_{\alpha i}\right)\right], \tag{5}$$

and the evolution equations for the O chemical potential are [31]:

$$\chi\frac{\partial\mu}{\partial t} = \nabla\cdot(\mathbf{M}\cdot\nabla\mu\,) - \sum_{\alpha i}\frac{\partial\rho}{\partial\eta_{\alpha i}}\frac{\partial\eta_{\alpha i}}{\partial t}, \tag{6}$$

where $\mathbf{M}$ is the O mobility tensor, which is explained in a later section, and $\chi = \frac{\partial\rho}{\partial\mu}$ is the susceptibility coefficient.

*Stress coupling*

Mechanical equilibrium requires that the stress state should satisfy:

$$\boldsymbol{\nabla}\cdot\boldsymbol{\sigma} = \mathbf{0}. \tag{7}$$

We use the Khachaturyan scheme (KHS) to obtain stress and elastic energy density values. In this framework, the stress tensor, $\boldsymbol{\sigma}$, is computed from Hooke's law relating the stress to the elastic strain, $\boldsymbol{\epsilon}^{\mathbf{e}}$, via the elasticity tensor $\mathbf{C}$:

$$\begin{aligned}\boldsymbol{\sigma} &= \mathbf{C}\colon\boldsymbol{\epsilon}^{\mathbf{e}}\\ &= \mathbf{C}\colon(\boldsymbol{\epsilon} - \boldsymbol{\epsilon}^{*}),\end{aligned} \tag{8}$$

where we have expanded the elastic strain into the total strain $\boldsymbol{\epsilon} = 1/2(\boldsymbol{\nabla}\mathbf{u} + \boldsymbol{\nabla}\mathbf{u}^{\mathrm{T}})$ and eigenstrain, $\boldsymbol{\epsilon}^{*}$. In the KHS interpolation method, the eigenstrain and the elasticity tensor, $\mathbf{C}$, are obtained from interpolations amongst the various phases as:

$$\begin{aligned}\mathbf{C} &= \sum_{\boldsymbol{\alpha}}\mathbf{h}^{\boldsymbol{\alpha}}\mathbf{C}^{\boldsymbol{\alpha}}\\ \boldsymbol{\epsilon}^{*} &= \sum_{\boldsymbol{\alpha}}\mathbf{h}^{\boldsymbol{\alpha}}\boldsymbol{\epsilon}^{*(\boldsymbol{\alpha})}.\end{aligned} \tag{9}$$

The elastic free energy density is calculated as:

$$f_{elastic} = \frac{1}{2}\boldsymbol{\sigma}\colon\boldsymbol{\epsilon}^{\mathbf{e}}. \tag{10}$$

The above expression is fed into the free energy functional (1) and the driving forces for the evolution of the order parameters are derived from it.

The eigenstrains for the monoclinic phase are calculated based on the monoclinic to tetragonal transformation while using the Zr phase as reference. We assume here that there is complete incoherency between the oxide phases and the zirconium substrate and complete coherency between the monoclinic grains. The transformation strain between the monoclinic and tetragonal phase is calculated using the minimum strain formulation in OptLat [37]. Strain variants are identified using crystal symmetry, and the ones chosen in the simulations are those that will produce the minimal strain in the x-direction of the simulation coordinate system, reducing the in-plane stress. Additional strain in the y-y direction has been added to make the volume change associated with the eigenstrains (their trace) consistent with the Pillings-Bedworth ratio of each phase. The two dominant variants for the monoclinic phase in the 2D simulations are given by:

$$\boldsymbol{\epsilon}_{v1}^{*(\mathbf{m})} = \begin{bmatrix} 0.0175 & 0.06 & 0 \\ 0.06 & 0.24 & 0 \\ 0 & 0 & 0.0159 \end{bmatrix}, \boldsymbol{\epsilon}_{v2}^{*(\mathbf{m})} = \begin{bmatrix} 0.0175 & -0.06 & 0 \\ -0.06 & 0.24 & 0 \\ 0 & 0 & 0.0159 \end{bmatrix}.$$

For the t-$ZrO_2$ we use

$$\boldsymbol{\epsilon}^{*(\mathbf{t})} = \begin{bmatrix} 0 & 0 & 0 \\ 0 & 0.18 & 0 \\ 0 & 0 & 0 \end{bmatrix}.$$

Due to the lack of information about the orientation relationship (and coherency) between the monoclinic grains and the h-ZrO phase we simply halve the value of the monoclinic eigenstrains due to there being half the oxygen content:

$$\boldsymbol{\epsilon}_{v1}^{*(\mathbf{h})} = \begin{bmatrix} 0.008175 & 0.03 & 0 \\ 0.03 & 0.12 & 0 \\ 0 & 0 & 0.007159 \end{bmatrix}, \boldsymbol{\epsilon}_{v2}^{*(\mathbf{h})} = \begin{bmatrix} 0.008175 & -0.03 & 0 \\ -0.03 & 0.12 & 0 \\ 0 & 0 & 0.007159 \end{bmatrix}.$$

This can be viewed as providing some level of coherency between the h-ZrO phase and the reference Zr phase. We note here that in the case where we have more than two order parameters in a phase, the order parameters used in each phase take one of their respective variant orientations above.  We end this subsection by noting that a different estimation for the orientation relationship between the monoclinic and h-ZrO phase and the resulting microstructural simulation is given in the Appendix

### *Treatment of O Diffusion*

The bulk contribution to O-diffusion includes contributions from various phases and is written in the form: $\sum_\alpha h^\alpha \mathbf{M}^\alpha$. Here $\mathbf{M}^\alpha = \chi^\alpha \mathbf{D}^\alpha$ is the mobility tensor in phase $\alpha$ (or the void), which is the product of the susceptibility coefficient and the diffusion tensor in that phase.

$$\mathbf{D}^\alpha = \mathbf{D_0} \exp\left(\frac{-\mathrm{Ea}}{\mathrm{kT}}\right) \tag{11}$$

Grain boundary diffusion can be a dominant transport mechanism for O and thus needs special attention. It is incorporated into the model by introducing the surface projection tensor $\mathbf{P}_S^{\alpha i,\beta j} = \mathbf{1} - \mathbf{n}^{\alpha i,\beta j} \otimes \mathbf{n}^{\alpha i,\beta j}$ which is comprised of the identity tensor and the outer product of the normal to the $\alpha i, \beta j$ interface. This normal vector is computed as $\mathbf{n}^{\alpha i,\beta j} = \nabla\phi^{\alpha i,\beta j}/\left|\nabla\phi^{\alpha i,\beta j}\right|$, where $\phi^{\alpha i,\beta j}$ is given by:

$$\nabla\phi^{\alpha i,\beta j} = \eta^{\alpha i}\nabla\eta^{\beta j} - \eta^{\beta j}\nabla\eta^{\alpha i}. \tag{12}$$

The contribution to the mobility from grain/interphase boundaries is given by: $\sum_{\alpha i,\beta j} M_S^{\alpha i,\beta j}\mathbf{P}_S^{\alpha i,\beta j}\left|\nabla\phi^{\alpha i,\beta j}\right|$, where the $\left|\nabla\phi^{\alpha i,\beta j}\right|$ term ensures that the total flux through an interface is independent of the artificial diffuse interface width. The final form for the mobility is given by:

$$\mathbf{M} = \left[\sum_{\alpha} h^{\alpha}\mathbf{M}^{\alpha} + \sum_{\alpha i,\beta j} M_S^{\alpha i,\beta j}\mathbf{P}_S^{\alpha i,\beta j}\left|\nabla\phi^{\alpha i,\beta j}\right|\right]. \tag{13}$$

*Homogeneous & Heterogeneous Grain Nucleation*

The present model employs a discrete stochastic nucleation algorithm in which nucleation events are treated as probabilistic processes. Classical nucleation theory provides the theoretical foundation for calculating both homogeneous and heterogeneous nucleation rates. Once a nucleation rate is computed, a stochastic sampling approach determines when and where nucleation occurs.

Based on classical nucleation theory (CNT) [38] the homogeneous nucleation rate is calculated as:

$$J^* = ZN\beta^* \exp\left(-\frac{\Delta G^*}{kT}\right), \tag{14}$$

where $Z$ is the Zeldovich factor (i.e., the probability that a nucleus at the top of the energy barrier will go on to form the new phase, rather than dissolve), $\beta^*$ is the rate of atom attachment to a critical nucleus, $N$ is is the number of nucleation sites, and $\Delta G^*$ is is the nucleation barrier evaluated at the critical nucleation radius. Formulas for the Zeldovich factor, Z, and the attachment rate, $\beta^*$, can be found in [35]. The free energy change is given by [35]:

$$\Delta G(R, c) = V(R)(\Delta f_{chem}^{\alpha}(c) + \Delta g_e) + A(R)\sigma_{h\alpha}. \tag{15}$$

In the above expression, $\Delta f_{chem}^{\alpha}$ is the Helmholtz chemical free energy density change, $\Delta g_e$ is the elastic free energy density change due to the volume change resulting from the introduction of the nucleus into the matrix [38], $\sigma_{h\alpha}$ is the surface energy density of the interface between the matrix and nucleus, $V(R)$ is the effective volume of the nucleus, and $A(R)$ is the surface area of the interface between the matrix and nucleus.

At the critical radius, where $\Delta G(R, c)$ is maximized with respect to R, the nucleation barrier, $\Delta G^*$, and the critical nucleus size $R^*$, is given by:

$$\Delta G^*_{hom}(R^*, c) = \frac{16\pi}{3}\frac{\sigma^3_{h\alpha}}{\left(\Delta f^{\alpha}_{chem}(c) + \Delta g_e\right)^2}$$
$$R^* = -\frac{2\sigma_{h\alpha}}{\Delta f^{\alpha}_{chem}(c) + \Delta g_e}. \quad (16)$$

We further annotate $\Delta G^*$ with a subscript "hom" to denote that it is the homogeneous nucleation barrier. Due to its high surface energy penalty, homogeneous nucleation tends to be highly improbable in solids compared to heterogeneous nucleation [39]. We consider the special case of a nucleus forming in a lenticular shape, as this is the minimum energy shape for a nucleus forming at a planar interface between two isotropic phases of dissimilar surface energy. In this widely considered case, it follows that the heterogeneous nucleation barrier can be related to the homogenous nucleation barrier in the form [40]:

$$\Delta G^*_{het} = \Delta G^*_{hom}\left[f\{\psi\} + \frac{\sin^3\psi}{\sin^3\psi'}f\{\psi'\}\right],$$

where $f\{\varphi\} \equiv \frac{2-3\cos\varphi+\cos^3\varphi}{4}$, and the angles can be found from a balance of surface tensions at the contact line of the nuclei:

$$\cos\psi = \frac{\sigma^2_{\alpha\beta} + \sigma^2_{\beta\gamma} - \sigma^2_{\alpha\gamma}}{2\sigma_{\beta\gamma}\sigma_{\alpha\beta}}$$
$$\cos\psi' = \frac{\sigma^2_{\alpha\gamma} + \sigma^2_{\beta\gamma} - \sigma^2_{\alpha\beta}}{2\sigma_{\beta\gamma}\sigma_{\alpha\gamma}}.$$

Nuclei at the critical radius are too small to be resolved on meshes suitable for large-scale simulations with large timesteps. To address this issue, an effective nucleation rate is used. It equates the volume formation rate of critical nuclei over a timestep with that of an effective nucleus, giving the number of effective nuclei per timestep as: $n^{1/3} = R^*/R_{eff}$, where $R_{eff}$ is user-defined, typically ~3 times the smallest element size. The effective surface nucleation rate is then:

$$J_{eff} = J^*/n. \quad (17)$$

We must correct the Monte Carlo sampling procedure used in the simulation which necessitates a volumetric nucleation rate, i.e., we sample a random number, $r$, at each Gaussian integration point and nucleate if $r < J_V dVdt$. This correction is obtained by observing the total number of nuclei introduced in a time, $T$, from a nucleation rate, $J_S$ on a surface $S_{ij}$ between grain i and j:

$$N = \int_T \int_{S_{ij}} J_{S_{ij}} dS\, dt = \int_T \int_V J_S \delta_{S_{ij}} dV\, dt. \quad (18)$$

where the second equality comes from the property of the surface Dirac function, $\delta_{S_{ij}} = |\nabla \phi^{\alpha i, \beta j}|$. The above equations also provide the correct volumetric nucleation rate, i.e., $J_V = J_{S_{ij}} \delta_{S_{ij}}$. Then, in the surface Monte-Carlo sampling procedure we sample a random number, r, and nucleate when the following criterion is met:

$$r < J_S \delta_S dVdt \quad (19)$$

for a given timestep size, $dt$ and element volume $dV$. Once nucleation is triggered thermodynamically, the code introduces it by modifying the order parameter fields, altering grains that intersect with the new nucleus. The nucleating order parameter is chosen based on the minimum value at the grain centroid to avoid unphysical coalescence of adjacent grains represented by the same parameter.

## 3. Results

Our phase field model requires various material property values as input. We calculated the lattice parameters of the relevant phases to determine eigenstrains, elastic constants, O diffusion coefficients in each phase, surface energies of each interface for the interfacial energy density, nucleation rates, and chemical free energies of each phase as a function of O concentration.

### *Structural and thermodynamic properties*

Experimental micrographs indicate that the most relevant $ZrO_x$ phases for pre-transition oxide formation are hcp α-Zr, hexagonal h-ZrO, and monoclinic m-$ZrO_2$ [20]. The computed lattice parameters and densities of α– and ω-Zr, h-ZrO, t-$ZrO_2$ and m-$ZrO_2$ phases are given in Table 3-1, together with experimental data, showing that the computed structural data are in good agreement with experiments.

*Table 3-1. Experimental and computed lattice parameters of the main phases of the Zr-O system. The computations were performed in the present work using DFT.*

| Zr phase | *a* (Å) | *b* (Å) | *c* (Å) | $\alpha$ | $\beta$ | $\gamma$ | Density ($g/cm^3$) |
|---|---|---|---|---|---|---|---|
| α-Zr (exp) [41] | 3.23 | 3.23 | 5.15 | 90 | 90 | 120 | 6.52 |
| α-Zr (comp) | 3.23 | 3.23 | 5.17 | 90 | 90 | 120 | 6.47 |
| **ω -Zr (comp)** | 5.05 | 5.05 | 3.15 | 90 | 90 | 120 | 6.55 |
| **α-Zr(O-sat)** | 5.65 | 5.65 | 5.22 | 90 | 90 | 120 | 6.66 |
| **h-ZrO (comp)** | 5.31 | 5.31 | 3.20 | 90 | 90 | 120 | 6.83 |
| **m-$ZrO_2$ (exp) [42]** | 5.15 | 5.21 | 5.32 | 90 | 99.2 | 90 | 5.82 |
| **m-$ZrO_2$ (comp)** | 5.19 | 5.25 | 5.38 | 90 | 99.6 | 90 | 5.66 |
| **t-$ZrO_2$ (exp) [43]** | 3.60 | 3.60 | 5.18 | 90 | 90 | 90 | 6.10 |
| **t-$ZrO_2$ (comp)** | 3.62 | 3.62 | 5.28 | 90 | 90 | 90 | 5.91 |

The present computations indicate the following stages in the oxidation of Zr metal, as illustrated in Figure 3-1. Starting with hcp $\alpha$-Zr, O atoms are dissolved exothermally with increasing O concentrations. Up to an O:Zr ratio of 0.33, $\alpha$-Zr remains the most stable phase. The upper limit is $Zr_3O$. The stability of the $\alpha$-Zr phase defines Stage 1 of oxidation. In stage 2, with increasing O concentration, the h-ZrO phase with the underlying $\omega$-Zr structure begins to form up to the stoichiometric composition of ZrO. Further addition of O leads to the formation of $ZrO_2$ in stage 3.

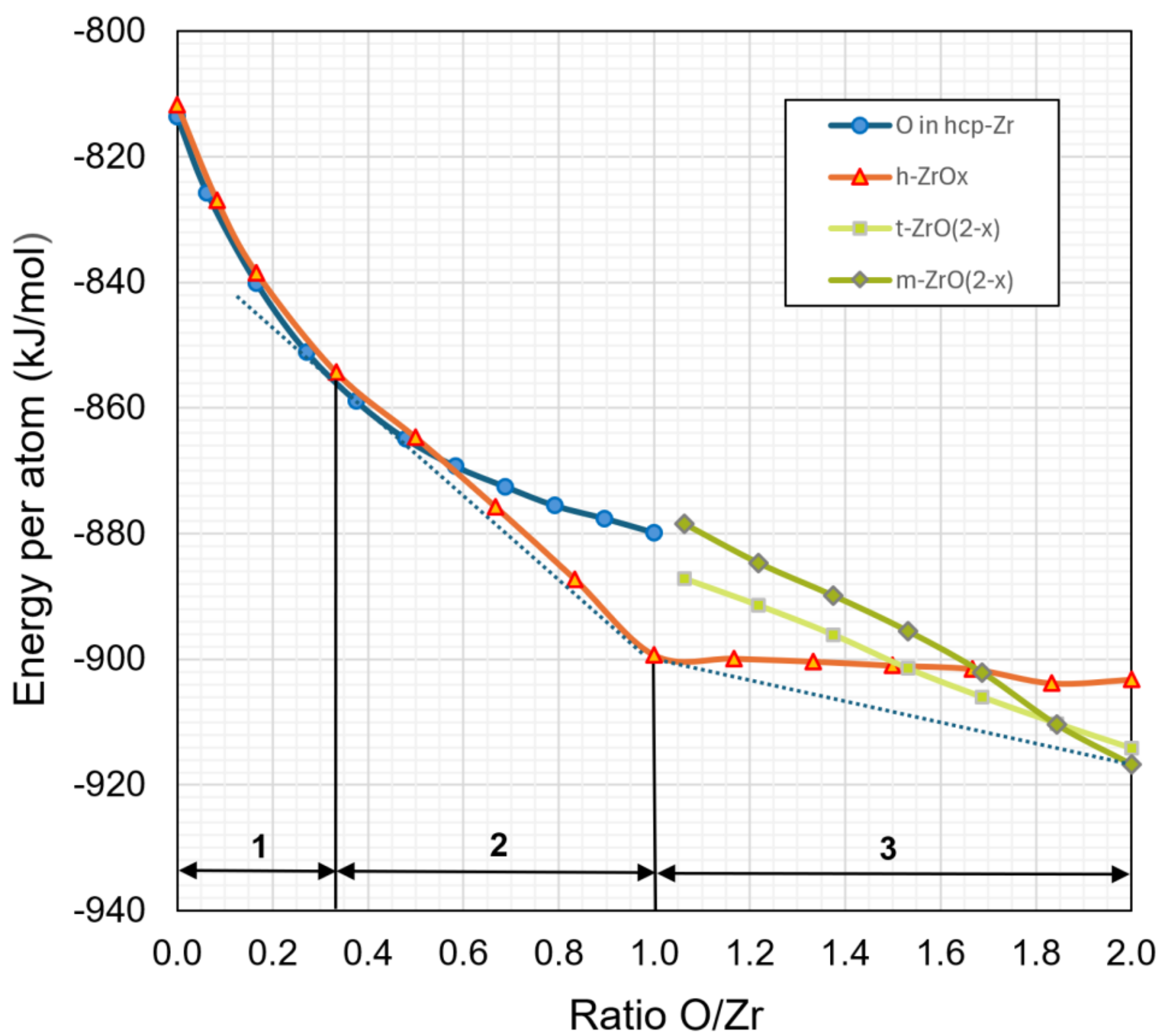

Figure 3-1. Total energies of Zr-O phases as a function of the O:Zr ratio, indicating a progression through three distinct oxidation stages, namely (1) the solution of O in hcp-

Zr metal until saturation (near $ZrO_{0.33}$), (2) the appearance of a h-ZrO phase, and (3) the formation of $ZrO_2$. The plotted energies include potential and kinetic terms obtained from molecular dynamics simulations in an NPT ensemble at 300 K using a DFT-PBE trained machine-learned potential.

For 1.0 < O:Zr < 2.0, at temperatures below the monoclinic to tetragonal transition temperature of 1400 K, and at ambient pressure, a thermodynamic equilibrium would exist between stoichiometric h-ZrO and m-$ZrO_2$. However, given that O vacancies and compressive stress stabilize the tetragonal phase, noting the small energy difference between the monoclinic and tetragonal $ZrO_2$ phases, and considering the complex dynamics of nucleation, it is quite likely that also the t-$ZrO_2$ phase forms in the transformation from the monoxide to the dioxide.

### *Elastic properties of $ZrO_x$*

Computed elastic coefficients and moduli are given in Table 3-2. Notably, the h-ZrO phase has the highest bulk and shear moduli. Monoclinic m-$ZrO_2$ has somewhat higher elastic moduli than tetragonal t-$ZrO_2$.

*Table 3-2. Calculated elastic coefficients of $\alpha$-Zr and Zr-O phases (GPa). B is the bulk modulus, S is the shear modulus, and E is the Young's modulus.*

| | $C_{11}$ | $C_{12}$ | $C_{13}$ | $C_{33}$ | $C_{44}$ | $C_{55}$ | $C_{66}$ | B | S | E |
|---|---|---|---|---|---|---|---|---|---|---|
| $\alpha$-Zr | 163 | 46 | 64 | 177 | 29 | | | 95 | 45 | 117 |
| $\alpha$-Zr(O-sat) | 198 | 109 | 91 | 230 | 75 | | | 135 | 61 | 160 |
| h-ZrO | 331 | 99 | 141 | 336 | 126 | | | 196 | 114 | 286 |
| t-$ZrO_2$ | 344 | 218 | 62 | 267 | 13 | | 151 | 183 | 77 | 203 |

| | $C_{11}$ | $C_{12}$ | $C_{13}$ | $C_{15}$ | $C_{22}$ | $C_{23}$ | $C_{25}$ | $C_{33}$ | $C_{35}$ | $C_{44}$ | $C_{46}$ | $C_{55}$ | $C_{66}$ | B | S | E |
|---|---|---|---|---|---|---|---|---|---|---|---|---|---|---|---|---|
| m-$ZrO_2$ | 318 | 150 | 111 | 39 | 365 | 133 | -2 | 282 | -8 | 75 | -6 | 81 | 120 | 195 | 94 | 243 |

### *Grain and interphase boundary energies*

Grain and interphase boundaries are modeled using polycrystalline models with two $ZrO_x$ phases. Such a model contains six to eight grains of each phase and has a size of 100x100x100 Å. Simulations of polycrystalline models are performed using the MLP. The volume and atomic positions are first relaxed. The grain boundary regions in the resulting models are typically 0.5 to 1 nm thick. Many O atoms in the grain boundary region have a lower atomic coordination than in the interior of the grains. Figure 3-2 shows an example of a computational model with $\alpha$-Zr and h-ZrO grains.

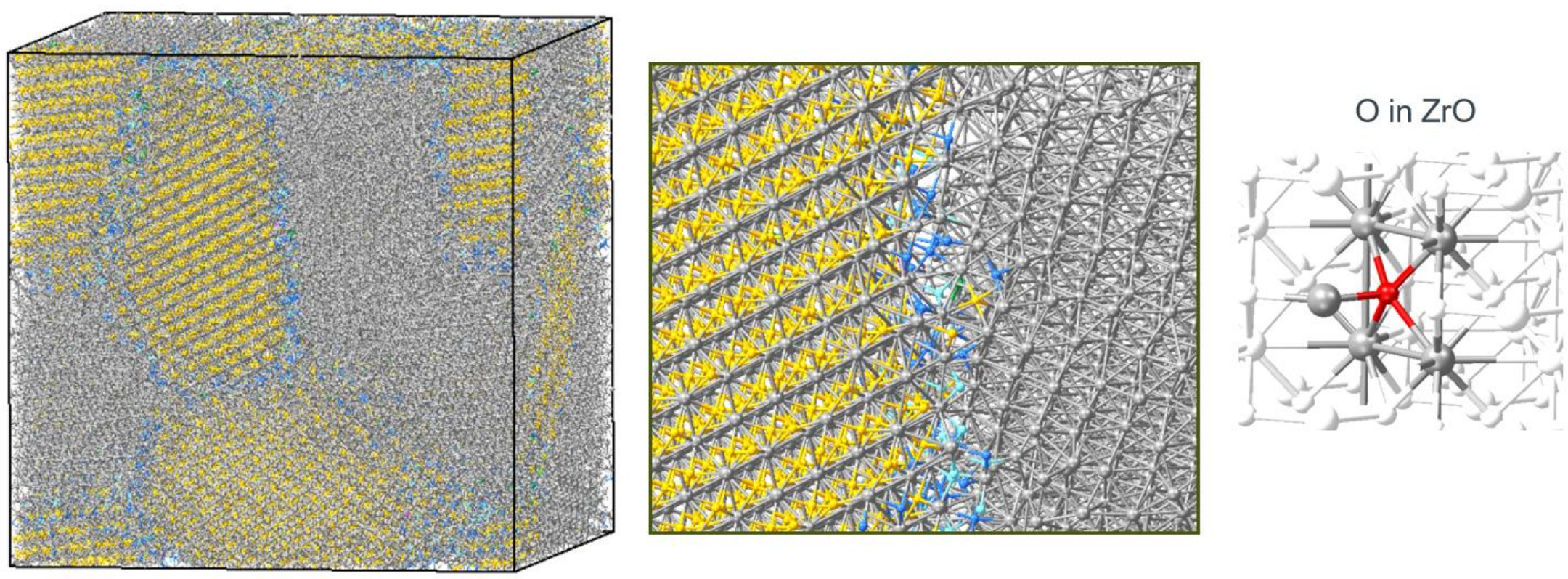

Figure 3-2. (Left) Polycrystalline Zr-ZrO model with O atoms color-coded according to their coordination number: blue indicates lower coordination, yellow indicates highest coordination. (Middle) Many O atoms in the grain boundary have a smaller coordination number (three or four) compared to O in the interior of ZrO (five). (Right) The five-fold coordinated O atom in ZrO.

The polycrystalline models are used to compute grain boundary energies for interfaces between each pair of considered $ZrO_x$ phases ($\alpha$-Zr, h-ZrO, t-$ZrO_2$ and m-$ZrO_2$). Since grains in the polycrystalline models are randomly oriented, we interpret computed grain boundary energies as average values for all possible grain boundaries in the model. The results are given in Table 3-3.

*Table 3-3. Average grain boundary energies (J/m$^2$) computed using polycrystalline models.*

| | **$\alpha$-Zr** | **h-ZrO** | **t-$ZrO_2$** | **m-$ZrO_2$** |
|---|---|---|---|---|
| **$\alpha$-Zr** | 0.50 | 1.39 | 1.39 | 1.01 |
| **h-ZrO** | | 1.80 | 1.52 | 1.64 |
| **t-$ZrO_2$** | | | 1.79 | 2.24 |
| **m-$ZrO_2$** | | | | 2.28 |

In general, each of the surface energies in the above table can be misorientation-, inclination-, and atom-termination-dependent. The values obtained above represent averages over this vast parameter space. We leave the exploration of this parameter space and the effect of surface energy variations on nucleation rates for future work.

### *Diffusion of O*

Table 3-4 gives diffusion barriers and pre-factors for bulk diffusion of O in $ZrO_x$, obtained using kinetic Monte Carlo simulations based on DFT energies. Hexagonal ZrO (h-ZrO) has the highest diffusion barriers, followed by monoclinic m-$ZrO_2$, $\alpha$-Zr, and t-$ZrO_2$ in that order. The phases h-ZrO and m-$ZrO_2$ are significant barriers for bulk O diffusion. Tetragonal $ZrO_2$ has a relatively high O diffusivity (faster than measured m-$ZrO_2$ grain boundary diffusion) with pronounced anisotropy. Diffusion of interstitial O dissolved in $\alpha$-Zr is somewhat slower than O diffusion in m-$ZrO_2$ grain boundaries.

*Table 3-4. Diffusion pre-factors $D_0$ and diffusion barriers $E_b$ for O diffusion in ZrOx phases. The ZrOx phases are listed in the order of the slowest to fastest diffusion in the table.*

| Phase | Direction | $D_0$ ($m^2/s$) | $E_b$ (kJ/mol) |
|---|---|---|---|
| h-ZrO | Avg | $1.47\times10^{-6}$ | 242.6 |
| | x | $3.39\times10^{-6}$ | 310.1 |
| | y | $2.10\times10^{-6}$ | 304.7 |
| | z | $4.41\times10^{-6}$ | 242.6 |
| m-$ZrO_2$ | Avg | $1.32\times10^{-7}$ | 227.5 |
| | x | $6.51\times10^{-8}$ | 220.5 |
| | y | $4.31\times10^{-7}$ | 233.5 |
| | z | $3.34\times10^{-8}$ | 242.9 |
| $\alpha$-Zr | Avg | $3.90\times10^{-6}$ | 188.5 |
| | x | $2.89\times10^{-6}$ | 188.4 |
| | y | $4.02\times10^{-6}$ | 188.5 |
| | z | $3.79\times10^{-6}$ | 188.5 |
| t-$ZrO_2$ | Avg | $1.98\times10^{-6}$ | 157.1 |
| | x | $2.81\times10^{-6}$ | 156.8 |
| | y | $2.84\times10^{-6}$ | 156.8 |
| | z | $3.08\times10^{-6}$ | 183.7 |

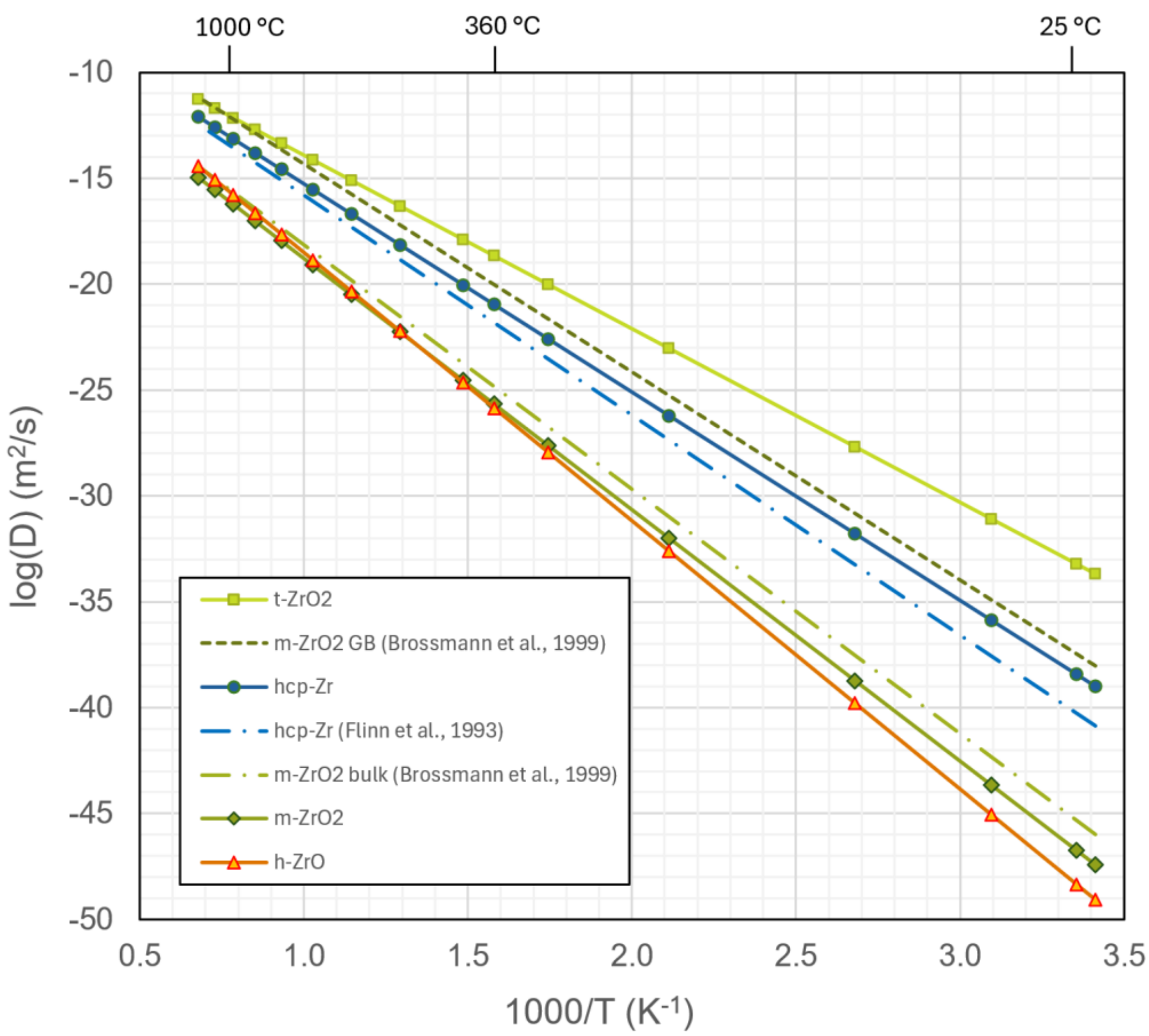

Figure 3-3. Diffusion of O in α-Zr and Zr oxides. The lines with markers are computed results, dashed and dashed-dotted lines are experimental values from Flint et al. [44] for O in α-Zr and from Brossmann et al. [45] for O in m-$ZrO_2$. Note the high diffusivity in t- $ZrO_2$ and the very low diffusivity in bulk m-$ZrO_2$ and h-ZrO. The diffusivity in grain boundaries (GBs) between m-$ZrO_2$ grains is very high compared with diffusion in bulk m- $ZrO_2$.

Figure 3-3 illustrates the large differences in diffusion coefficients between different phases in the Zr-O system. A comparison of the computed diffusivities with experimental results from Flinn et al. [44] and Brossmann et al. [45] as illustrated in Figure 3-3 shows good agreement, considering the large uncertainties in the experimental data as discussed in the aforementioned publications.

In the idealized limit of a continuous, defect-free layer of either h-ZrO or m-$ZrO_2$, the very low bulk diffusivities computed for these phases would severely limit oxygen transport and dramatically slow oxidation. While real oxide layers grown in pile inevitably contain defects, radiation damage, and other heterogeneities that provide additional transport pathways, this analysis underscores a robust conclusion: grain boundaries and grain boundary diffusion play a dominant role in the aqueous corrosion of Zr alloys under the conditions studied here.

### Oxide Microstructural Evolution

Next, we show the computed nucleation rates for various nucleation mechanisms, followed by our phase field simulations.

### Nucleation dynamics

Given both the chemical similarity of t-$ZrO_2$ and m-$ZrO_2$ along with the similarity of the interface energies (interchanging m-$ZrO_2$ and t-$ZrO_2$), the results of the nucleation analysis for the m-$ZrO_2$ phase are also used for the t-$ZrO_2$ phase.

Nucleation analysis: Heterogeneous nucleation of m-$ZrO_2$ grains at the Zr/m-$ZrO_2$ interface

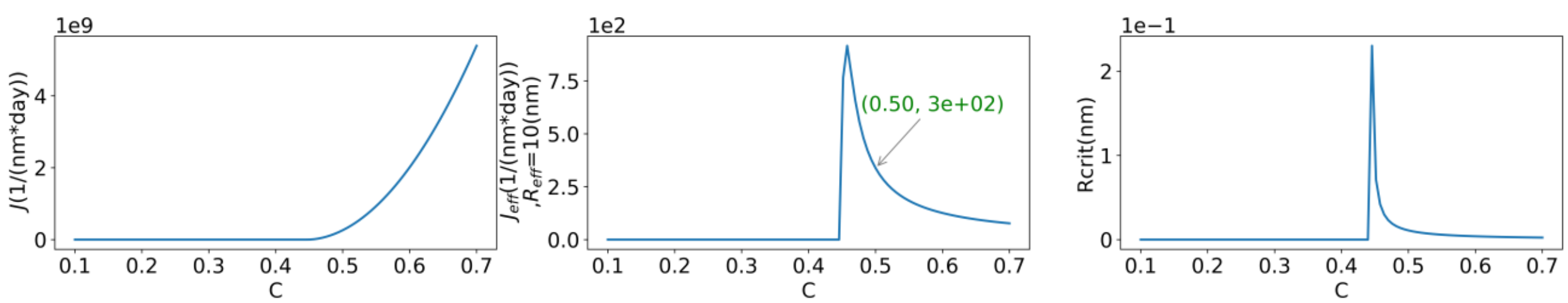

Figure 3-4. m-$ZrO_2$ grain nucleation rates at the Zr/m-$ZrO_2$ interface versus O atom fraction c (c = 0.667 corresponds to $ZrO_2$.) (left) Classical nucleation rate. (center) Effective nucleation rate for 10 nm nuclei. The marked value is used for the effective nucleation rate in the simulations. (right) Critical nucleus size showing composition dependence.

Figure 3-4 illustrates the nucleation behavior of m-$ZrO_2$ grains at the Zr/m-$ZrO_2$ interface as a function of composition. The classical nucleation rate, J in Figure 3-4 (left) shows a strong compositional dependence, increasing exponentially with higher O atom fractions. The effective nucleation rate, $J_{eff}$, for 10 nm nuclei, calculated using (17), exhibits different behavior as shown in Figure 3-4 (center). $J_{eff}$ peaks at an O atom fraction of 0.45 with a maximum value around 8e2 $(nm^3 \cdot day)^{-1}$ before declining at higher O atom fractions. The value at a 0.5 mole fraction, corresponding to 3e2 $(nm^3 \cdot day)^{-1}$ for the effective nucleation rate, is marked to denote the value used in the subsequent simulations. Figure 3-4 (right) demonstrates that the critical nucleus size, $R_{crit}$, determined from (16), also displays composition dependence, peaking at 0.25 nm near

c = 0.45, which closely corresponds to the peak in the effective nucleation rate, before decreasing at higher c values.

*Nucleation analysis: Heterogeneous nucleation of m-$ZrO_2$ grains at the h-ZrO/m-$ZrO_2$ interface*

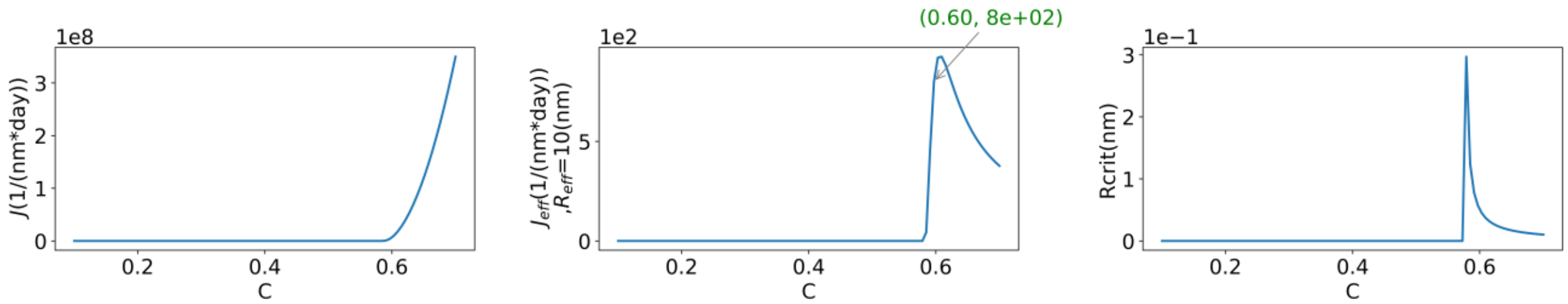

Figure 3-5. m-$ZrO_2$ grain nucleation results at the h-ZrO/m-$ZrO_2$ interface versus O atom fraction c. (left) Classical nucleation rate. (center) Effective nucleation rate for 10 nm nuclei at a concentration of 0.6, the value at the interface, is marked, denoting what is used for the effective nucleation rate in the preceding simulations. (right) Critical nucleus size showing composition dependence.

A similar analysis was performed for heterogeneous nucleation of m-$ZrO_2$ grains at the h-ZrO/m-$ZrO_2$ interface in Figure 3-5. Notably, the effective nucleation rate here is similar to the effective nucleation rate of Zr/m-$ZrO_2$. The effective nucleation rate used in the current simulations is also depicted on the middle plot. This nucleation mechanism is the primary mechanism for the continued development of oxide grains and texture formation within the oxide.

*Nucleation analysis: Heterogeneous nucleation of h-ZrO grains at Zr/m-$ZrO_2$ interface*

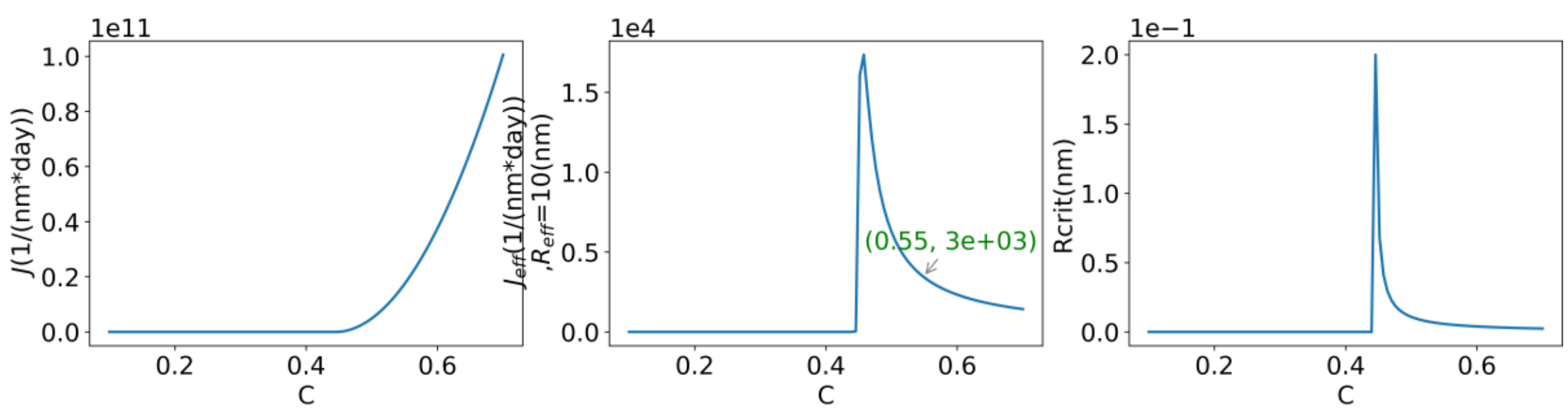

Figure 3-6. h-ZrO grain nucleation results at the Zr/m-$ZrO_2$ interface versus composition O atom fraction c. (left) Classical nucleation rate. (middle) Effective nucleation rate for 10 nm nuclei with a concentration of 0.55 is marked, denoting what is used for the effective nucleation rate in the preceding simulations based on the concentration seen at the interface. (right) Critical nucleus size showing composition dependence.

Figure 3-6 shows that h-ZrO grain nucleation at the Zr/m-$ZrO_2$ interface, r-behaves similarly as at other interfaces except at different rates. The classical nucleation rate (left) remains minimal until an O concentration of 0.5, then increases sharply to $10^{11}$ $(nm^3 \cdot day)^{-1}$. The effective nucleation rate (middle) peaks at an O concentration of 0.45, reaching 1.6e4 $(nm^3 \cdot day)^{-1}$. Critical nucleus size (right) spikes at an O concentration of 0.4 to a critical radius value of 0.2 nm. This nucleation mechanism is likely responsible for generating the observed h-ZrO grain structure, given its high rate.

*Nucleation summary*

The schematic in Figure 3-7 illustrates the three main nucleation sites in the $ZrO_x$ system. The layered structure shows m-$ZrO_2$ (top), h-ZrO (middle), and $\alpha$-Zr (bottom). Yellow circles at the Zr/h-ZrO interface represent h-ZrO nucleation sites. Nucleation of m-$ZrO_2$ grains (green circles) at an h-ZrO/m-$ZrO_2$ interface is also highly likely and is judged to be the main mechanism for continued texture development of the monoclinic phase. We note that the nucleation rates were calculated assuming isotropic eigenstrains because analytical expressions for the stored elastic energy are known for this case. A discussion of the effects of relaxation of this assumption and its implications on the nucleation rates derived is reserved for future work.

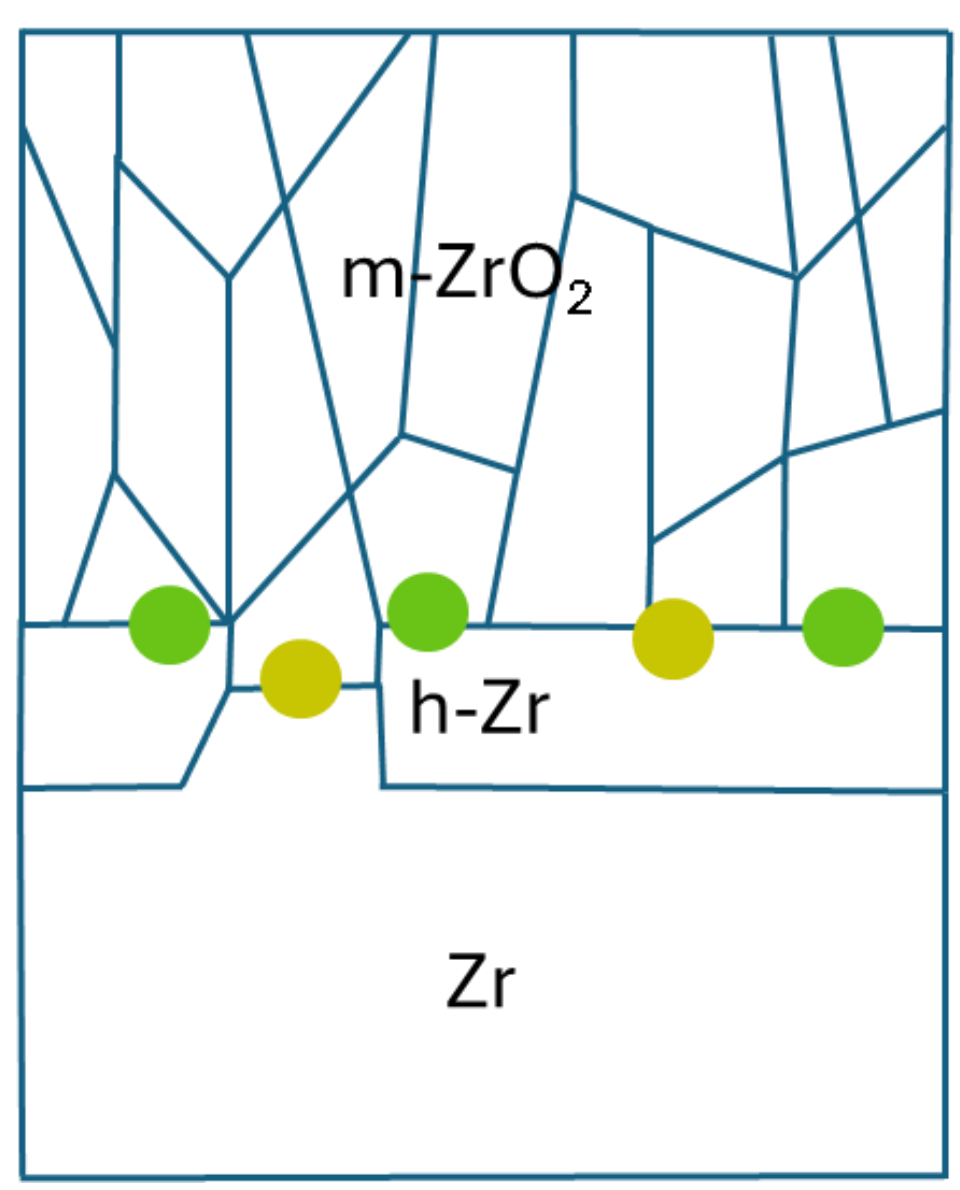

Figure 3-7. Schematic of the possible nucleation mechanisms for the Zr, h-ZrO, and m-$ZrO_2$ system. h-ZrO nucleates at the metal oxide interface (yellow circles) and monoclinic $ZrO_2$ grains nucleate at the Zr/h-ZrO interface (green circles).

We end this section by directing the reader to the Appendix where a sensitivity analysis is performed to study how changing the diffusivity, surface energy, and lattice mismatch impact the nucleation rates in this section.

## Phase-Field Simulations of Microstructural Evolution in Zirconium Oxide Systems

In the following simulations, we demonstrate the behavior of the derived phase-field model under various conditions. For simplicity, we assume here that the O chemical potential is constant at the surface. The first simulation includes only m-$ZrO_2$ and Zr phases to simplify the analysis, showing how columnar grain structures develop with corresponding stress field distributions. Finally, we incorporate the t-$ZrO_2$ and h-ZrO phases to analyze their impact on the grain morphology, particularly near the metal-oxide interface where characterization studies confirm the presence of h-ZrO. All simulations are in 2D and run at 600K with a thin starting layer of m-$ZrO_2$ grains on top of a Zr substrate.

### Stress distribution in monoclinic phase

The simulated microstructure and corresponding stress field distributions for a system with monoclinic and zirconium phases after 25 days is shown in Figure 3-8.

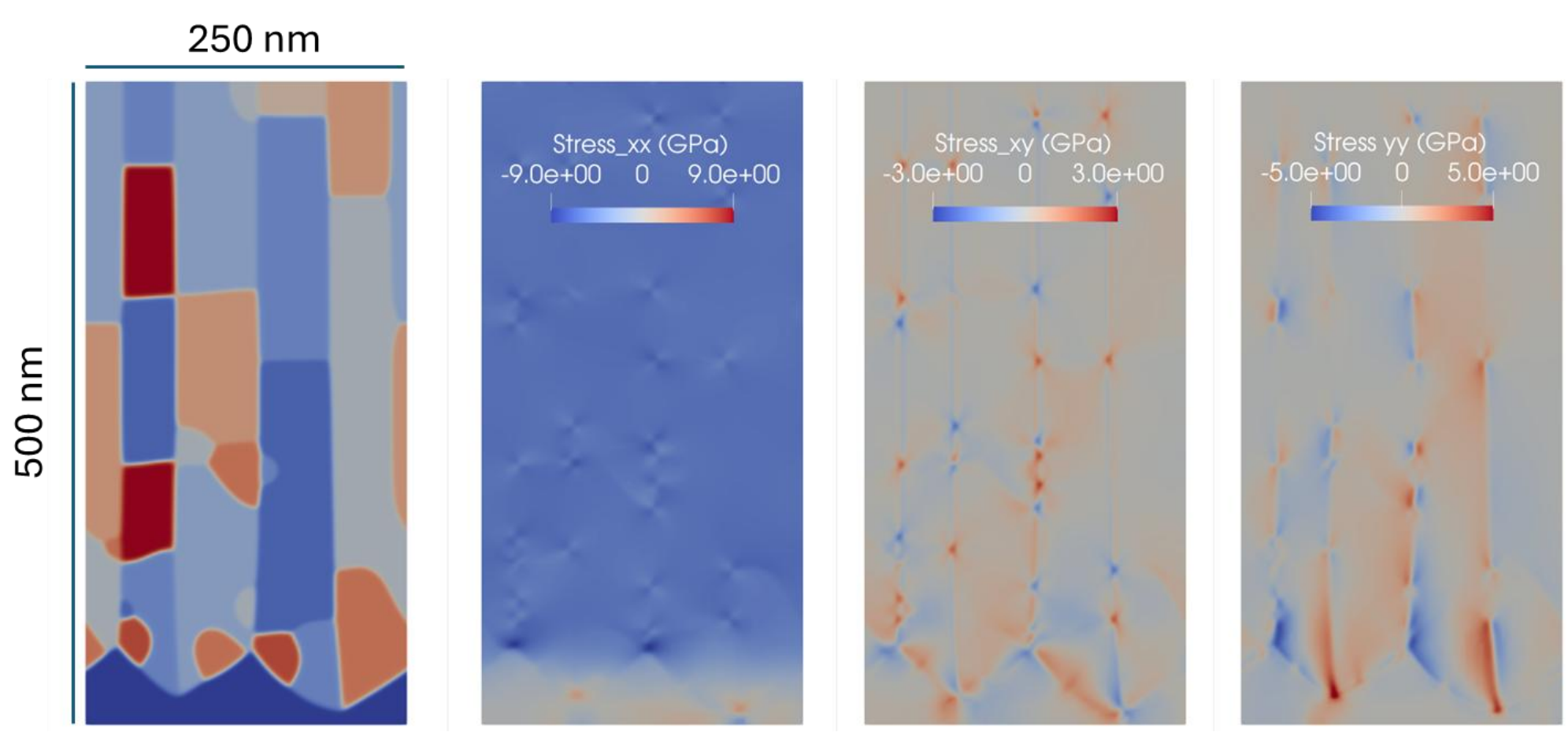

Figure 3-8. From left to right: Simulated microstructure with predominantly columnar grains; Cauchy stress xx-component; xy-component; and yy-component, showing predominantly compressive stresses with fluctuations near grain boundaries due to varying grain orientations.

The microstructure (far left) reveals predominantly columnar grains oriented in the vertical direction. The stress field components exhibit several key characteristics. The oxide experiences predominantly compressive stress throughout the domain, as evidenced in the xx-component (middle left). Near grain boundaries, significant stress fluctuations occur due to differences in grain orientations. Notably, regions with relatively large tensile stresses appear in the yy-direction component of stress at grain boundary interfaces (normal to the average metal/oxide interface and parallel to the direction of mass transport), which could initiate horizontal crack formation. The xy-component (middle right) shows considerable shear stresses, particularly at grain boundaries, resulting from lattice mismatch between adjacent monoclinic strain variant orientations. These interfacial stress concentrations play a critical role in determining the stability of the oxide layer as they are the predominant driving forces acting in the oxide other than the capillary driving forces (akin to surface tension in fluids) resulting from surface energy minimizations. The simulations performed here do not include plasticity/creep mechanisms for reducing the stress, thus these driving forces should be viewed as an upper bound.

### Introduction of the h-ZrO and t-$ZrO_2$ phases

Next, we added both h-ZrO and t-$ZrO_2$ phases into the simulation to see how it impacts the overall microstructure in Figure 3-9. In this four-phase system, h-ZrO forms as an intermediate layer between the α-Zr substrate and the m-$ZrO_2$ oxide, consistent with experimental observations near the metal-oxide interface. The presence of h-ZrO alters the local stress states and grain morphologies, producing a more refined microstructure with smaller oxide grains. The h-ZrO layer acts as a structural bridge between the metallic substrate and the oxide, making it easier to accommodate lattice mismatch which can also be seen by a gradual transition in the pressure in part c) of the figure. The introduction of this phase seems to influence the subsequent growth of m-$ZrO_2$ grains above it where we can see more equiaxed grains near the undulation of the h- ZrO/m-$ZrO_2$ interface. We note that the first layer of m-$ZrO_2$ grains at this interface are freshly nucleated grains and have not had the opportunity to grow and are thus small and equiaxed, as is also found in experiment. Finally, the small presence of the t-$ZrO_2$ phase near the metal oxide interface is also observed in experiment.

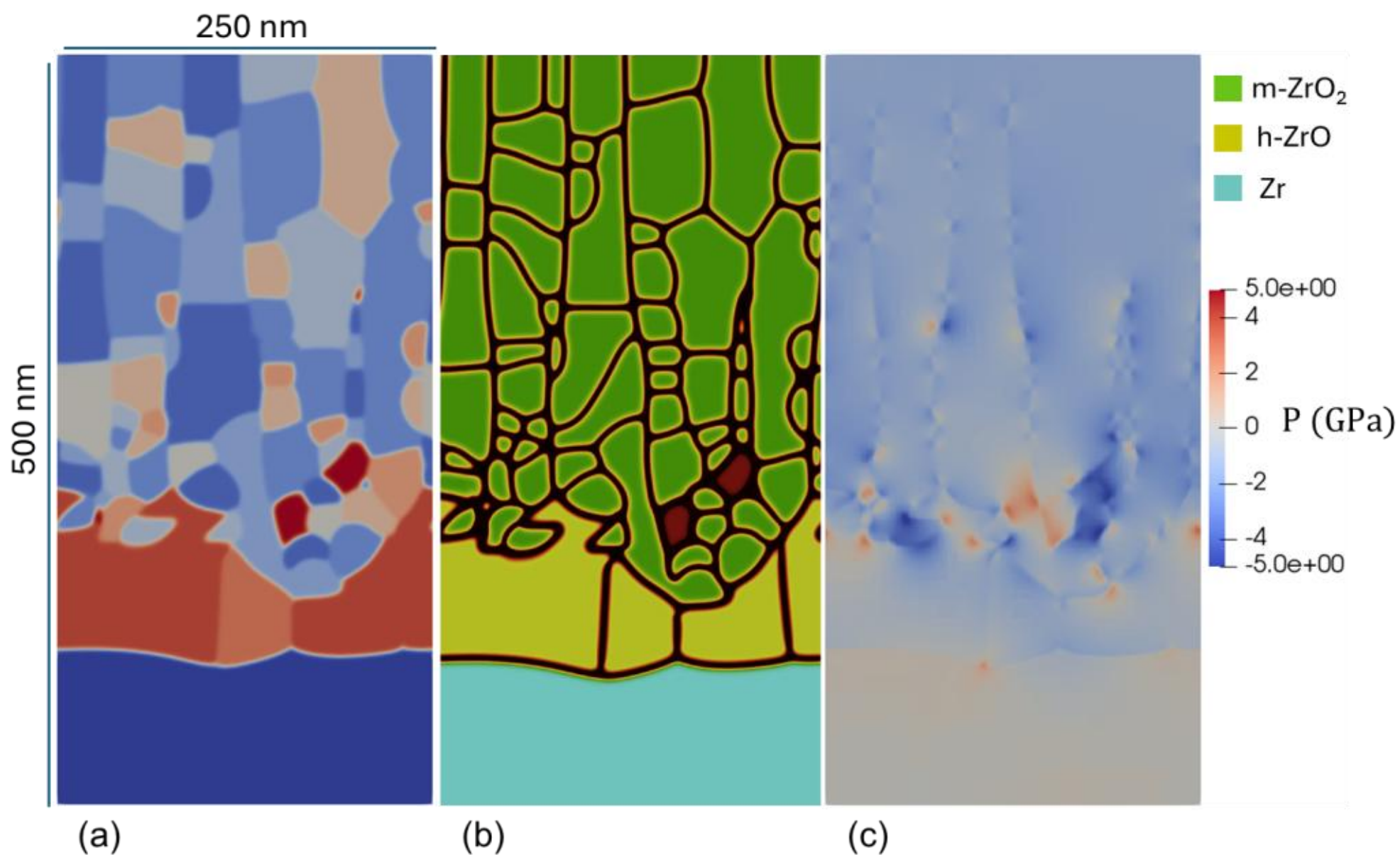

Figure 3-9. Phase-field simulation of oxide growth showing layered microstructure development. (a) Individual grain variants colored by crystallographic orientation within a 250×500 nm simulation domain. (b) Same microstructure colored by phase identity: monoclinic $ZrO_2$ (green) tetragonal $ZrO_2$ (red), hexagonal ZrO (yellow), and α-Zr substrate (light blue). (c) Pressure distribution highlighting the heterogeneous distribution of tensile and compressive stresses. The simulation illustrates the formation of m/t-$ZrO_2$ grains on top of h-ZrO grains on the metallic substrate.

### Microstructure characterization and validation

Statistical characterization of heterogeneous materials requires quantitative descriptors to distinguish geometric features. The cluster pair correlation function C2(r) measures the probability that two randomly selected points separated by distance r belong to the same connected cluster of a specific phase [46]. We compute this function for various simulations and compare it to the experimental function to quantitatively compare the microstructures.

*Monoclinic grain statistical analysis*

We start by analyzing the experimental images of the monoclinic grains taken and modified from [47]. Figure 3-10 shows the microstructural analysis of a large area (≈6750 nm × 2000 nm) containing mostly monoclinic grains. Color quantization (Figure a) is applied to reduce the color noise of the experimental data, followed by phase identification (Figure 3-10 b) that highlights monoclinic dioxide grains in yellow. The size-based grain mapping (Figure 3-10 c) distinguishes individual grains by their volume with yellow being the largest. The cluster distribution function (Figure 3-10 d) reveals an anisotropic pattern, confirming the directional nature of grain growth and a slight tilt in the data shown by highlighted regions in this plot.

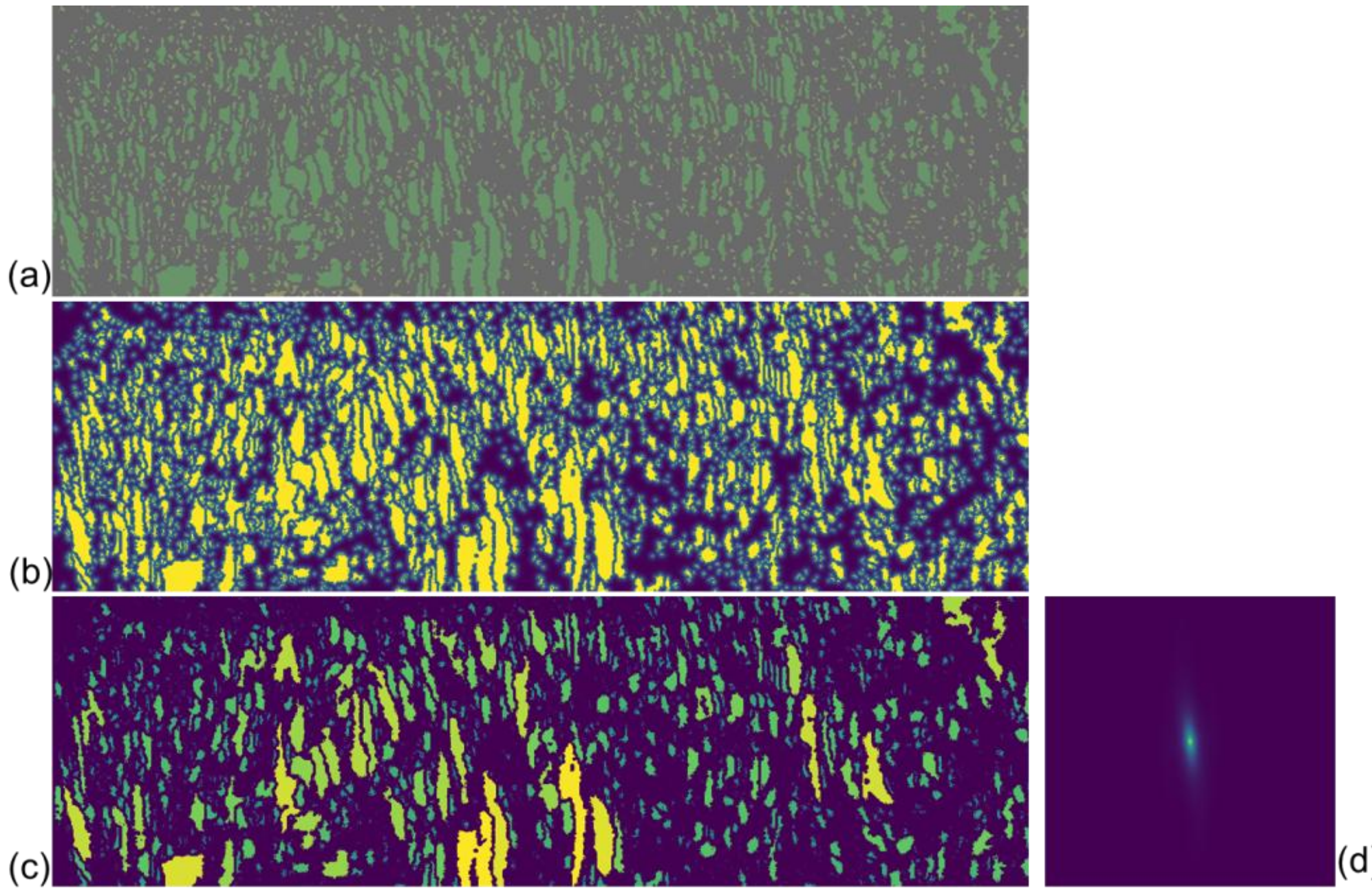

Figure 3-10 (a) Color quantization filter applied to the top portion of the cropped experimental microstructure image (≈ 6750 nm × 2000 nm); (b) Phase identification filter highlighting monoclinic dioxide grains; (c) Grains mapped and colored by their sizes; and (d) Cluster distribution function $C_2$ for the identified grain network.

The size distribution function of the grains in the larger sample is shown in Figure 3-11. The size distribution function is a plot of the number of grains whose volume is that of a circle with a given diameter. The figure only shows those grains whose diameters are greater than 20 nm to filter out some of the smaller grains which we interpret as image noise, showing a pronounced decrease in frequency as the diameter increases.

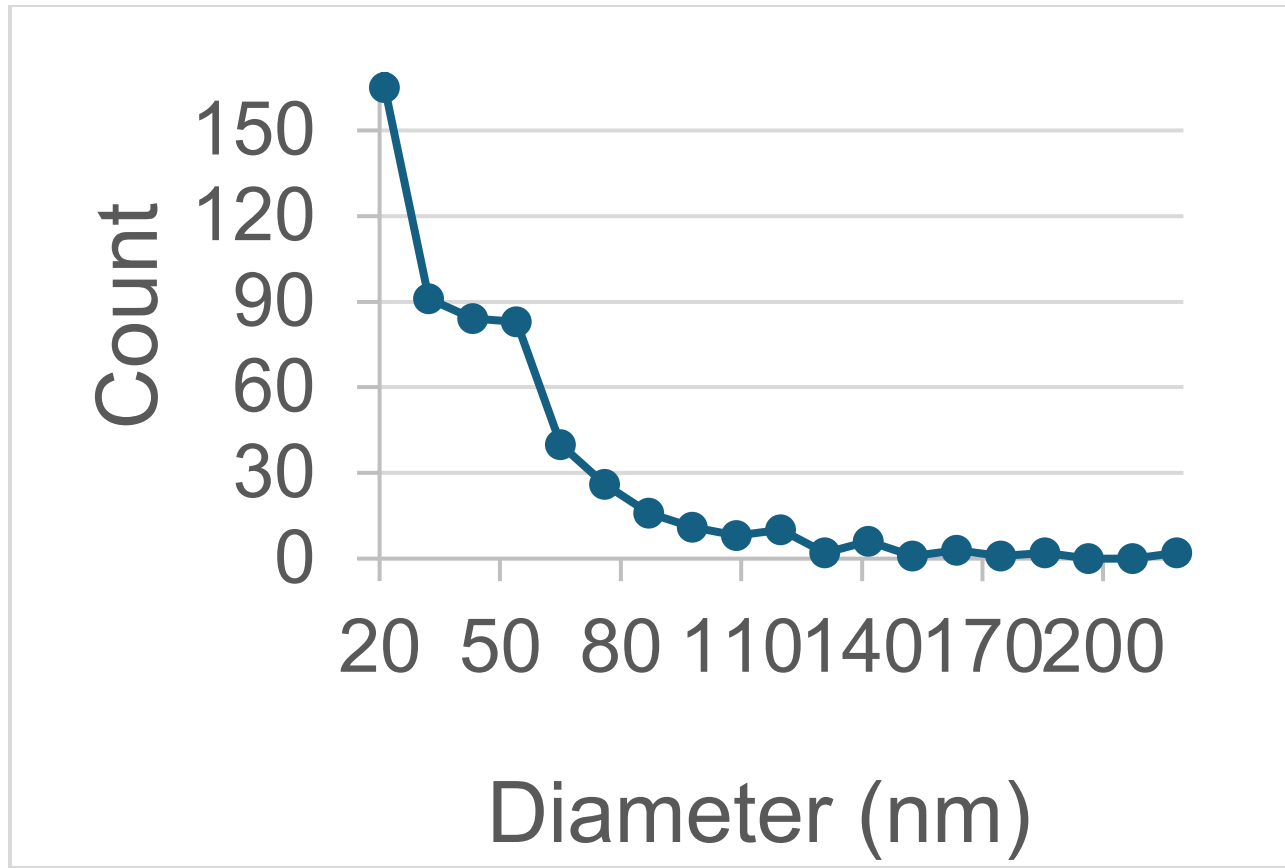

Figure 3-11 Size distribution function of effective circles with diameters greater than 20 nm, showing a rapid decline in frequency with increasing size.

The distribution exhibits two notable features: a prominent peak at the smallest measured diameter range (approximately 20-30 nm) and a secondary shoulder around 50 nm. The primary peak in the smaller-diameter range could be attributed to large numbers of small-grain structures and/or noisy experimental data in this size range. The secondary peak centered at 50 nm likely represents the mean grain size of the material, after accounting for image noise. The distribution shows a rapid decline after the 50 nm peak, with count values decreasing substantially as the diameter increases. By 100 nm, the frequency has diminished to near-zero levels, indicating that larger grains are rare in this material. The presence of sporadic counts at diameters beyond 100 nm may indicate potential data-quality issues caused by two neighboring grains being mistaken for a single large grain.

To quantify the columnar nature of the grains, ellipses are fitted to each grain using the algorithm in [48]. The algorithm determines the optimal ellipse parameters (center, axis lengths, and orientation) for each grain that preserve the centroid and second-order spatial moments of the original grain structure. The microstructure of the monoclinic grain structure of the large sample, with ellipses superimposed on individual grains, is shown in Figure 3-12.

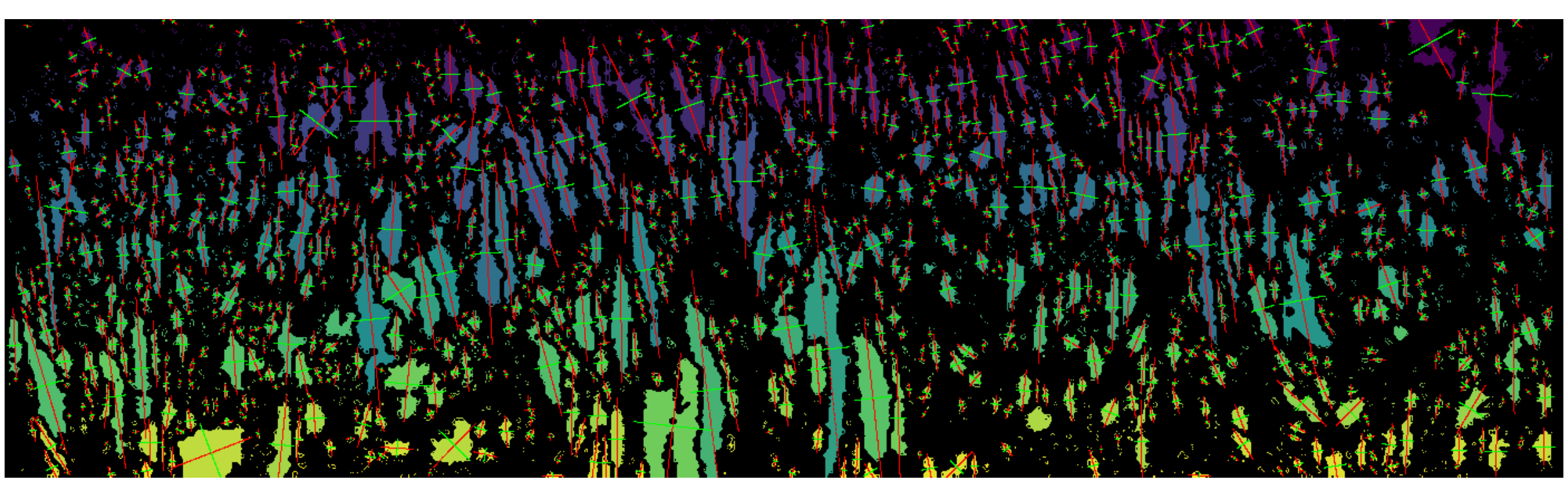

Figure 3-12. Image of the monoclinic microstructure with ellipses fitted to individual grains; red and green lines denote the major and minor axes, respectively. Average grain statistics: major axis length = 53.6 nm, minor axis length = 18.5 nm, aspect ratio = 2.8, and average orientation angle = 85.95°.

The microstructure shows elongated grains with a strong preferred orientation, as evidenced by the nearly vertical alignment of the red major axis lines, while the green minor axis lines indicate the minor axis of each fitted ellipse. Averaging over all the grains results in an average major axis length of 53.6 nm, a minor axis length of 18.5 nm, resulting in a relatively high aspect ratio of 2.8 and an average orientation angle of 85.95° (nearly vertical). In Figure 3-13, we plot the orientation distribution for completeness. It shows precisely how preferential the vertical orientation of the grains is.

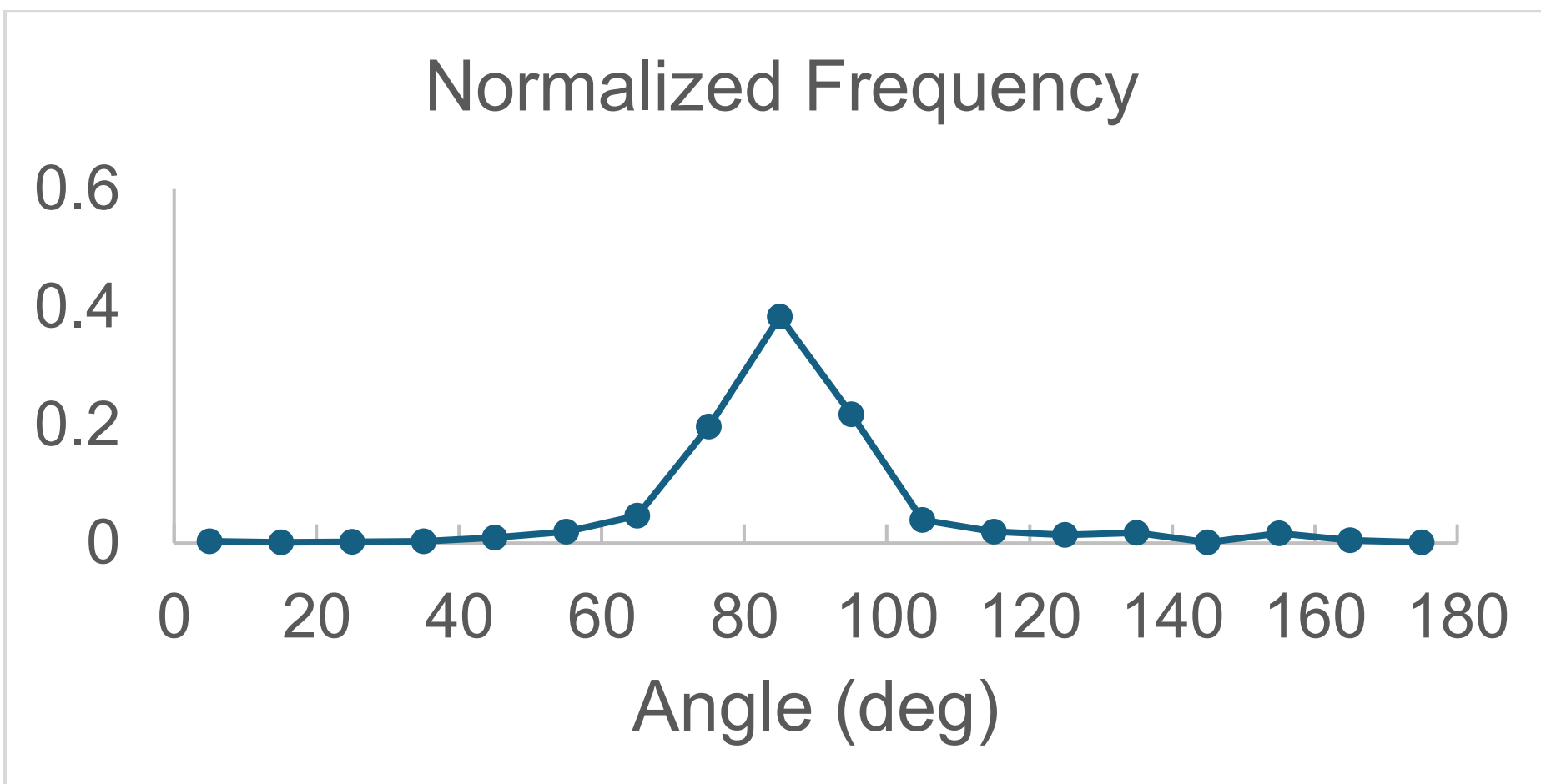

Figure 3-13. Angular distribution of monoclinic grains, showing a strong peak near 85.95°, indicating a predominant grain alignment close to the vertical direction.

Having computed the statistics of the experimental image, we can now compare that to the simulated data. Below, we show a visual comparison between simulated and experimental data.

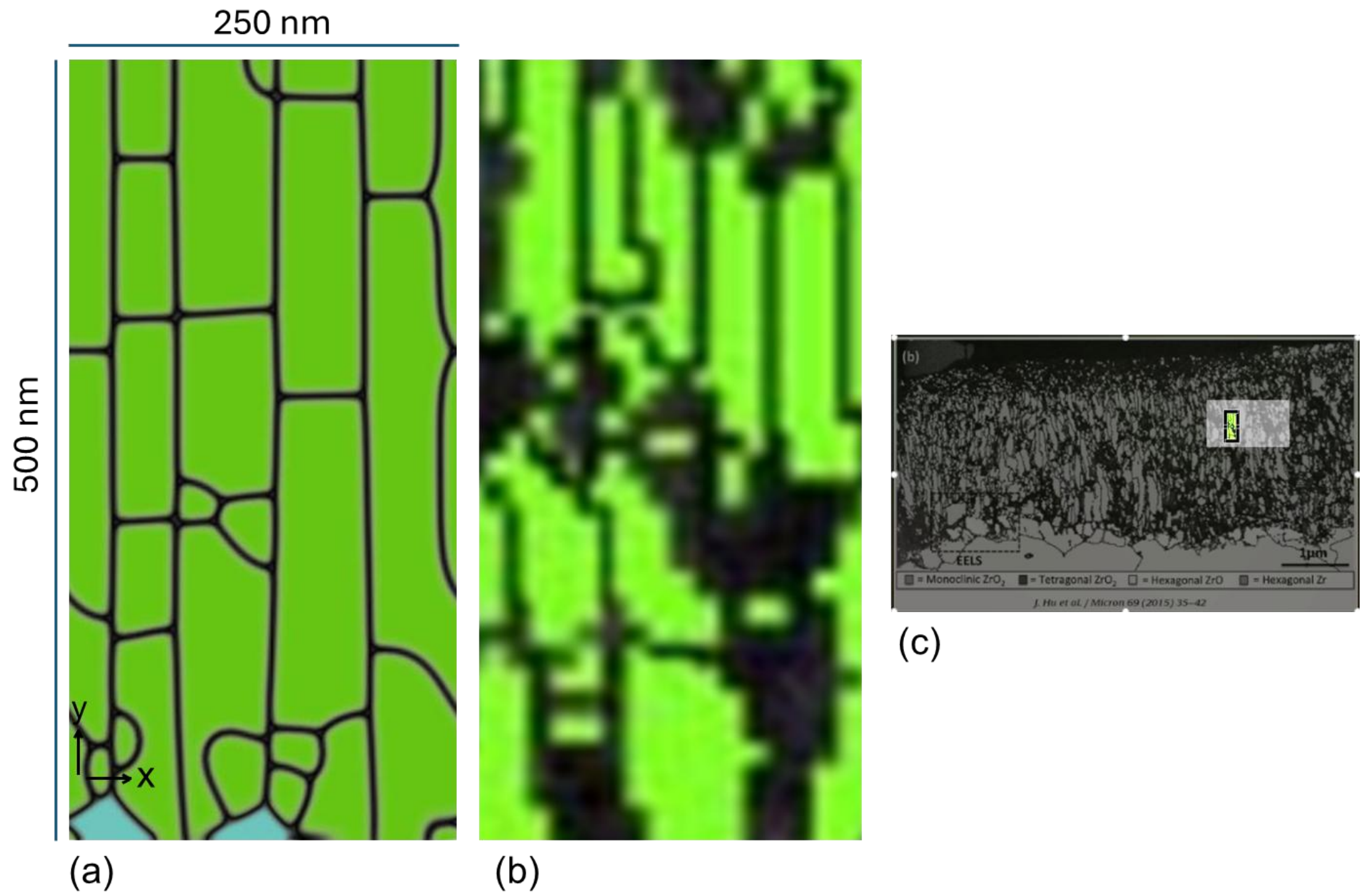

Figure 3-14. Visual microstructure comparison. (a) Simulated microstructure (250 nm × 500 nm) with color-coded grains. (b) Corresponding experimental microstructure visualized in green. (c) Location of (b) within the larger experimental sample. The strong visual correspondence between simulation and experiment validates the computational model.

Figure 3-14. compares simulated and experimental microstructures at the same scale (250 nm × 500 nm). The grain distribution patterns in both images are substantially similar, validating the computational approach used in the simulation. Both microstructures exhibit columnar grain structures with high aspect ratios, as evidenced by the vertically elongated grain formations. These columnar structures appear consistently in both the simulation and experimental sample.

A key difference is the absence of tilting in the simulated structure compared to the visible tilt in the experimental microstructure (shown on the larger-scale image). This discrepancy likely results from factors not incorporated into the simulation. Three potential mechanisms could reproduce the observed ~4° tilt: (i) a macroscopic stress gradient operating across domains larger than the 250 nm × 500 nm simulation volume, which would bias grain growth directions; (ii) crystallographic texture in the α-Zr substrate that templates oxide nucleation with systematic angular offsets; or (iii) epitaxial or semi-coherent orientation relationships between the metal substrate and initial oxide nuclei that establish preferential growth directions propagating through the film thickness. Incorporating these mechanisms requires either larger simulation domains or explicit treatment of crystallographic orientation correlations between substrate and oxide phases. We did not attempt preliminary tests of these extensions given the substantial computational and theoretical development required, but these represent clear directions for model refinement.

Next, we analyze the cluster pair correlation function below for this simulation to quantify the similarities of the microstructure.

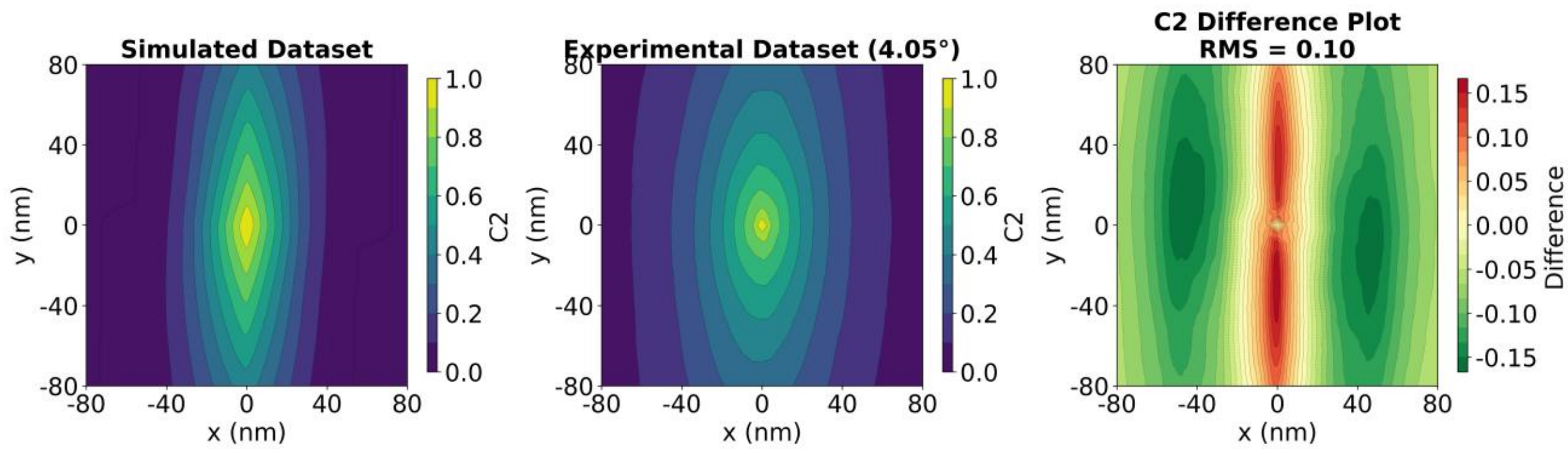

Figure 3-15. Comparison of anisotropic cluster distribution functions for monoclinic grains. (a) Simulated C2 (unitless) function showing elliptical distribution pattern. (b) Experimental C2 function (rotated 4.05°) with similar elongated structure. (c) Difference plot showing a) minus b) revealing greatest deviations between experimental and simulated data. The color scale indicates the magnitude of variation in the central region, with an RMS value of 0.098.

Figure 3-15 compares anisotropic cluster distribution functions ($C_2$) for monoclinic grains between simulation and experiment. The figure shows strong similarities in the overall distribution patterns, exhibiting pronounced elliptical structures. The simulated $C_2$ function (Figure 3-15 a) displays a well-defined elliptical distribution pattern that closely matches the experimental $C_2$ function (Figure 3-15 b), which has been rotated 4.05° to align the data in the vertical direction. Both distributions reflect the underlying columnar grain structure of the material, as evidenced by the vertical elongation of the elliptical patterns. The difference plot (Figure 3-15 c) reveals two regions of discrepancy: the first occurs (yellow region) in the vertical region in the center of the distribution. These discrepancies may be attributed to the smaller size correlation features visible in the experimental data, which could result from measurement noise rather than actual grain differences. The second is at larger distances from the center. This may be a simulation domain size effect on the simulated structure due to grains being truncated via reaching the end of the simulation window. However, both differences are quite small with an RMS value of 0.098, as becomes evident in a broader visual comparison of the two structures in Figure 3-14..

*Hexagonal grain analysis*

Next, we analyze the experimental image near the metal oxide interface to gather statistics about the h-ZrO phase for comparison to our simulated data in the last section. Figure 3-16 presents a morphological analysis of the h-ZrO phase near the metal-oxide interface. The original micrograph (Figure 3-16 a) reveals a sparse distribution of h-ZrO grains across the 4,250 nm field of view, indicating this phase constitutes only a minor volume fraction of the microstructure. The color-quantized image (Figure 3-16 b) enhances phase contrast, while the size-coded representation (Figure 3-16 c) shows variation in grain volumes. The ellipse-fitted analysis (Figure 3-16 d) shows that h-ZrO grains are predominantly equiaxed rather than displaying the columnar structure observed in monoclinic grains. The major (green) and minor (red) axes show aspect ratios lower than those of the columnar monoclinic phase. The cluster distribution function (Figure 3-16 e) captures spatial correlation patterns, with the directional anisotropy indicated in (Figure 3-16 f) revealing a slight tilt from the horizontal orientation. The circular marker in Figure 3-16 f indicates the average grain size.

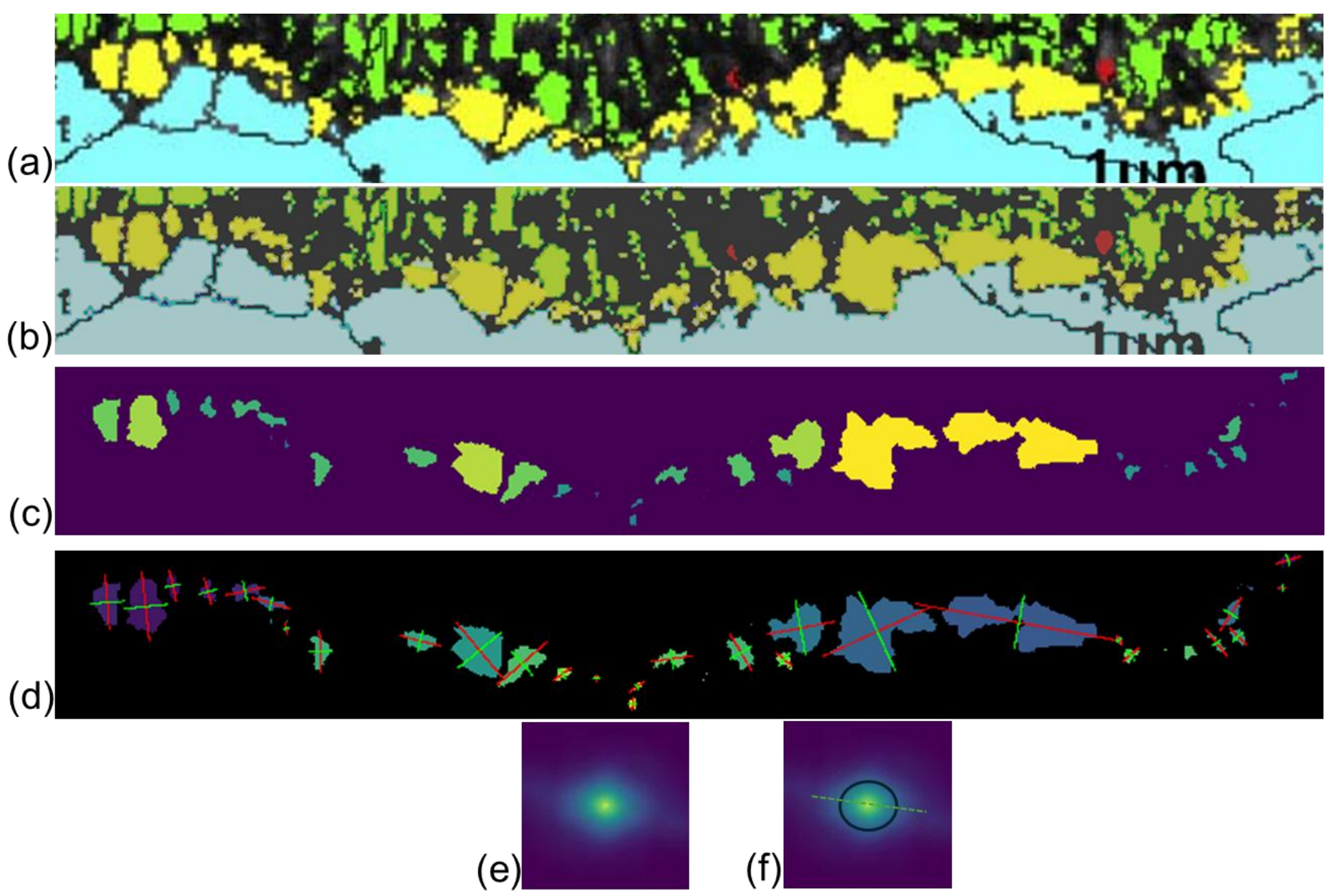

Figure 3-16. Morphological analysis of h-ZrO phase near metal-oxide interface. (a) Original experimental micrograph showing h-ZrO phase distribution (~4250 nm x 550

nm). (b) Color-quantized image with reduced noise for improved phase distinction. (c) Individual h-ZrO grains colored by size. (d) Grain shape analysis with fitted ellipses showing major (green) and minor (red) axes. (e) Cluster distribution function showing spatial correlation patterns with highly correlated regions in yellow and uncorrelated regions in dark blue. (f) The same distribution function in (e) with horizontal direction is highlighted to show anisotropy, and the circular marker indicates the average grain size.

Figure 3-17 and Figure 3-18 compare cluster distribution functions ($C_2$) between simulated and experimental datasets for the h-ZrO and m-$ZrO_2$ grains respectively. The simulated $C_2$ function in (Figure 3-17a) displays a radially symmetric distribution with concentric contours, closely matching the pattern observed in the experimental data (Figure 3-17b) at small distances from the center. Both simulated and experimental distributions exhibit similar central peaks and circular contour patterns, quantitatively confirming the more equiaxed nature of h-ZrO grains compared to monoclinic grains. The difference plot (Figure 3-17c) shows that the greatest deviations occur at scattered peripheral locations corresponding to anisotropy in the simulated data.

The similarity between simulation and experiment is notable, though we acknowledge that the h-ZrO phase constitutes only a minor volume fraction of the microstructure, yielding limited grain counts for robust statistical comparison. The RMS value of 0.18 should therefore be interpreted cautiously. Nevertheless, the agreement in fundamental morphological character, equiaxed h-ZrO grains versus columnar m-$ZrO_2$, supports the conclusion that the model correctly captures the essential nucleation and growth physics at the metal-oxide interface. Additional experimental characterization across multiple fields of view would strengthen this validation. The $C_2$ function for monoclinic grains closely resembles the experimental function, though with a slightly higher RMS value compared to the monoclinic-only simulation shown in Figure 3-15.

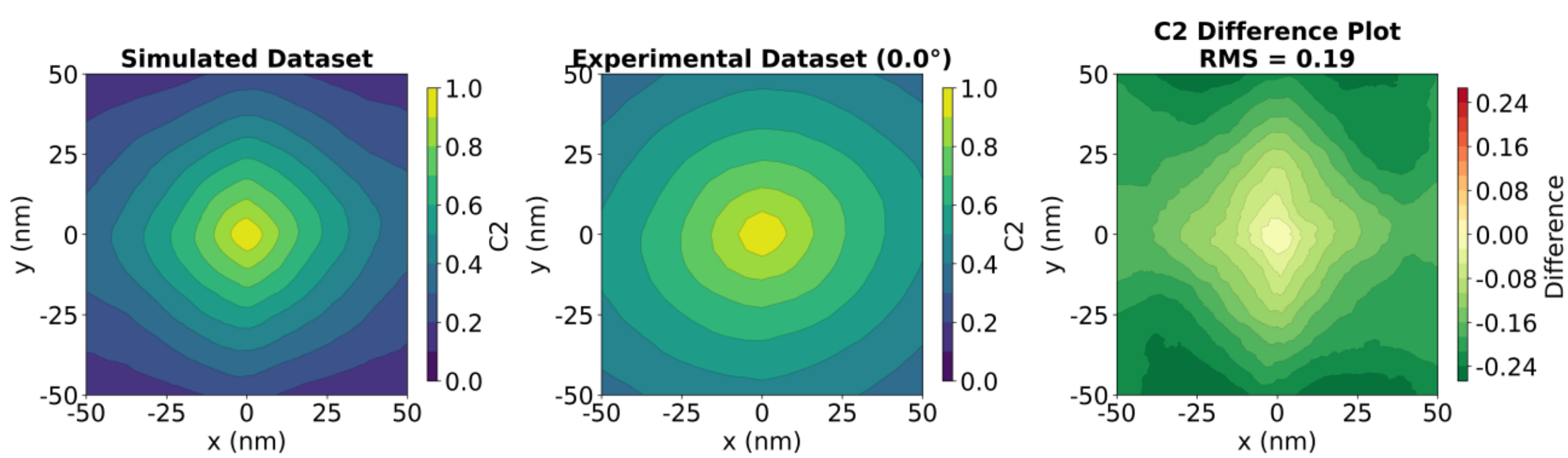

Figure 3-17. Comparison of cluster distribution functions in simulated and experimental datasets of h-ZrO grains. (a) Contour plot of simulated $C_2$ function showing mostly radially symmetric distribution (100 nm × 100 nm). (b) Corresponding experimental $C_2$ function

with similar central peak and concentric patterns. (c) Difference plot highlighting regions of greatest deviation between simulation and experiment, with scale indicating magnitude of discrepancy with an RMS value of 0.18.

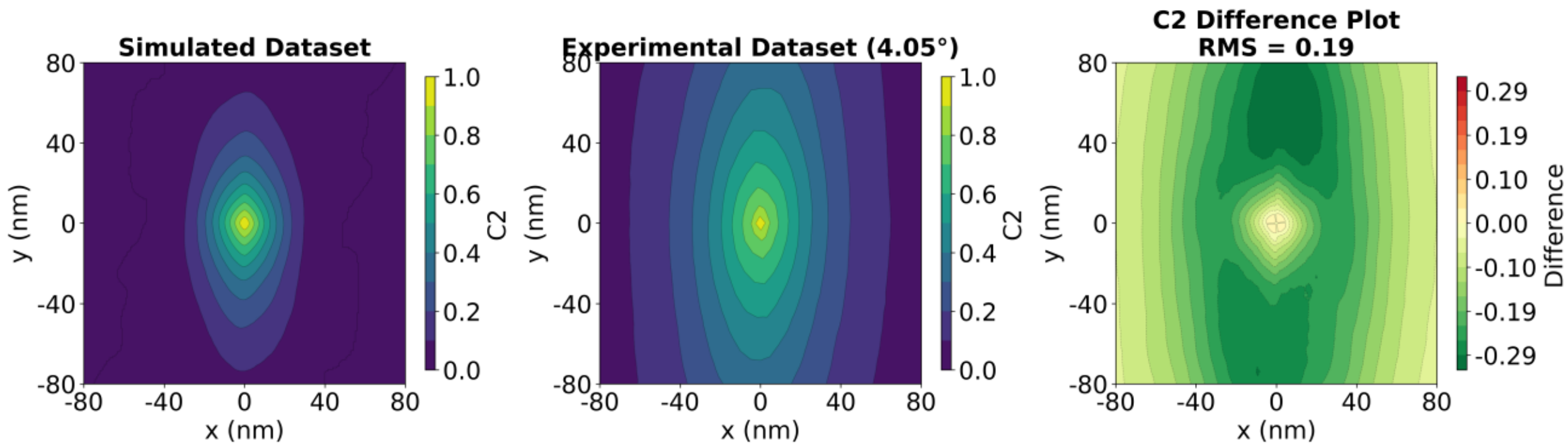

Figure 3-18. Cluster pair correlation function $C_2$ for monoclinic grains in the four-phase simulation (α-Zr, h-ZrO, t-$ZrO_2$, m-$ZrO_2$). (a) Simulated $C_2$ function exhibiting vertical elongation characteristic of columnar grain morphology. (b) Experimental $C_2$ function corrected for sample tilt (4.05° rotation). (c) Difference between simulated and experimental distributions (RMS = 0.19). The color scale indicates the magnitude of variation in the central region, with an RMS value of 0.19.

Below in Figure 3-19, we compare the simulated microstructures with (b) and without (a) the h-ZrO and t-$ZrO_2$ phases being present. This is plotted against the experimental micrograph in part (c). The simulation with the h-ZrO phase present shows elongated monoclinic grains but with lower aspect ratios than those of the simulation without h-ZrO present. This can also be seen by the larger differing $C_2$ functions in the vertical direction in Figure 3-18 c compared to Figure 3-15 c. We suspect that this is due to the change in the stress state near the metal oxide interface, impacting the elastic forces between neighboring monoclinic grains. It is also possible that the orientation relationship between the h-ZrO and monoclinic grains, and in turn the eigenstrains defined in Section 3, may need further study. Since data are scarce on this interphase boundary and its preferred orientations, we leave this problem for future study.

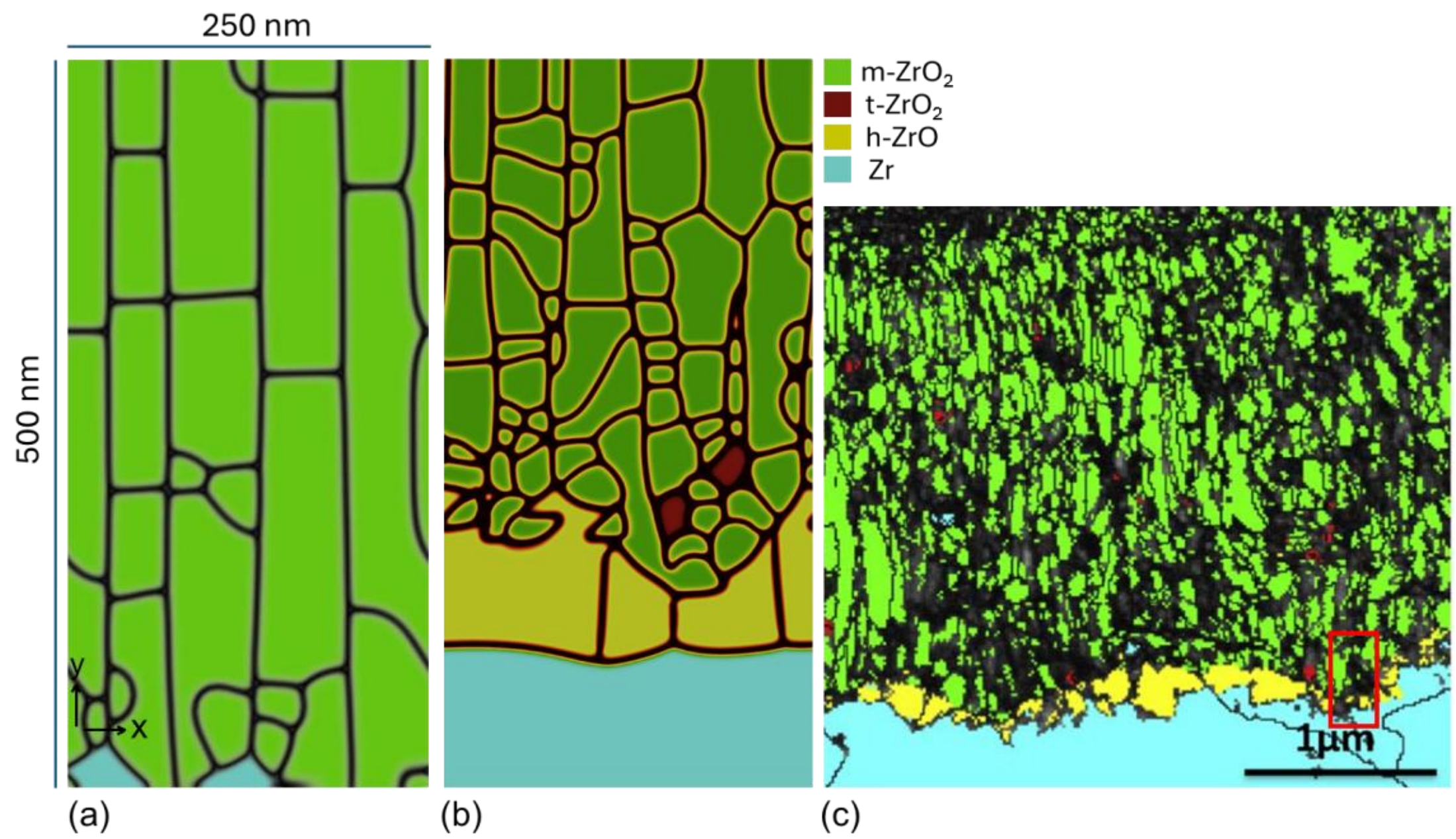

Figure 3-19. Visual comparison of microstructure without (a) and with the h-ZrO phase (b) next to a cropped version of the experimental micrograph (c). A red box is placed in (c) to show the size of the simulated domains in parts (a) and (b) to make visual comparisons easier.

We also note here that the addition on the h-ZrO phase at the metal oxide interface seems to have produced a longer wavelength perturbation of the dioxide interface compared to the simulation with only monoclinic and zirconium phases. Another possible future extension of this work would be to study this.

# 4. Discussion

The above findings suggest that quasi-1D models of Zr aqueous corrosion, while computationally efficient, must incorporate microstructure-informed corrections to capture these grain-scale effects.

As with the stabilized h-ZrO layer, our simulations show that the competition between grain boundary diffusion and grain nucleation similarly creates a metal-oxide interface with a naturally undulating topology. In 3D, this tendency would develop into the so-called "cauliflower" interface texture widely observed in experiments. In turn, the undulating nature of the metal-oxide interface creates regions of oxide with tensile stress components, despite the overall compressive state of the oxide film. Localized tensile stresses at interface undulations and grain boundary triple junctions can initiate mode I vertical (i.e., transverse) cracks, while steep stress gradients at the undulatory interface create bending moments that produce tensile stresses on the convex side of undulations, providing nucleation sites for mode I lateral cracks. The vertical spacing of lateral cracks could correlate with the scale of undulation development, with cracking occurring when undulation amplitude reaches a critical value sufficient to exceed the oxide fracture strength. This would continue as the oxide grows leading to a vertical spacing of horizontal cracks. The current model, which does not include crack nucleation or growth, identifies the most likely mechanism driving both vertical and horizontal cracking. Clearly including crack modelling would be a logical extension of the current work.

The simulations presented here were performed in 2D due to computational constraints; each simulation was run on a 12-core Intel i7 workstation, and transitioning to 3D at equivalent resolution would necessitate substantially increased HPC resources, but is feasible with the current tool. This 2D treatment introduces some limitations: in 3D, the grain boundary area to volume ratio increases, providing additional diffusion pathways and allowing access to more energetically favorable strain variants. However, we do not expect the columnar grain morphology to change moving to 3D simulations. The columnar structure emerges from heterogeneous nucleation at the metal-oxide interface, where local oxygen concentrations govern nucleation rates. Since oxygen transport remains the rate-limiting process regardless of dimensionality, the preferential nucleation and growth of oxide grains perpendicular to the interface should persist in 3D, though the spatial distribution of grain orientations may differ due to the additional stress relaxation introduced by the additional strain variants. These additional variants will also aid in producing undulations in the new in-plane direction of the metal oxide interface and are expected to increase the undulatory “cauliflower” nature of the metal-oxide interface. Once rate transition and cracks appear in the corrosion film, crack connectivity in 3D would also differ. However, the present work focused on pretransition oxidation microstructure.

## 5. Conclusions

The investigation reported herein has established a computational framework for $ZrO_x$ phase evolution during aqueous Zr corrosion, using as inputs only thermodynamic, transport, and elastic property values which were either measured or rigorously computed via first principles and atomistic methods. To this end, the dependences of the free energies of oxygen-saturated $\alpha$-Zr and three $ZrO_x$ phases on O:Zr ratio were accurately mapped, enabling microstructural simulations with dramatically improved fidelity. Elastic properties and oxygen diffusion coefficients were calculated for all these phases.

Our simulations successfully reproduced several critical features of observed pre-transition corrosion film microstructures. Specifically, they captured the microstructural evolution of a aqueous corrosion film on a Zr substrate from an equiaxed, primarily h-ZrO oxide layer adjacent to the metal/oxide interface into a columnar, primarily m-$ZrO_2$ grain structure. The specific crystallographic relationships between the oxide and substrate were not modeled, however. Systematic microstructural characterization performed in this work has revealed clusters of roughly equiaxed h-ZrO grains near the metal-oxide interface beneath columnar m-$ZrO_2$ grains with a mean aspect ratio of 2.8 and an orientation angle of 85.95°. Quantitative validation of our results against this experimental data using cluster pair distribution functions demonstrates good agreement, with RMS values of 0.098 for monoclinic grains and 0.18 for h-ZrO simulations, confirming the predictive capability of the current computational approach.

The current simulations elucidate the important role played by the hexagonal ZrO layer near the metal/oxide interface, and the transverse grain boundaries within it, in the resulting pretransition oxidation kinetics. In fact, our simulations indicate that a continuous, undefected layer of h-ZrO at the metal-oxide interface, were one to develop, would essentially shut down the corrosion process altogether and lead to a passivated oxide film of essentially constant thickness. Complete passivation does *not* occur in real corrosion films because of the presence of transverse grain boundaries and other discontinuities in the h-ZrO layer, that are continually being created by the heterogeneous nucleation process. However, the formation of partial, defected h-ZrO layer could play a role in depressing the pretransition aqueous corrosion rate on Zr surfaces from a parabolic to a cubic trend. Inasmuch as pretransition Zr aqueous corrosion kinetics in and out of pile are not markedly dissimilar, we expect that the formation of an h-ZrO layer close to the metal-oxide interface is important for pretransition corrosion behavior under irradiated conditions as well. We speculate that soluble alloying additions to Zr that would promote h-ZrO stability might further reduce aqueous corrosion rates and perhaps also delay rate transition.

**Acknowledgments**

The authors would like to thank Bruce Kammenzind (US Naval Nuclear Laboratory), Bob Comstock, Eric Mader, John Beale, and Aylin Kucuk (Electric Power Research Institute) for their valuable discussions and support throughout this work. Marthe Bideault (Materials Design) contributed to the calculation of diffusion in oxides.

Additional computational resources were provided by EPRI on their Apollo compute cluster, whose support was highly appreciated. We acknowledge the EuroHPC Joint Undertaking for awarding this project access to the EuroHPC supercomputer LUMI, hosted by CSC (Finland) and the LUMI consortium through a EuroHPC Regular Access call (project 465000610).

## 6. Appendix

### Sensitivity study of h-ZrO/m-ZrO2 orientation relationship

The eigenstrains for the hexagonal phase in this section are calculated based on the hexagonal to monoclinic transformation while using the Zr phase as reference. We again utilize OptiLat and follow the same procedure defined in the text to obtain the coherency eigenstrain and are left with a single variant:

$$\boldsymbol{\epsilon}_{\mathrm{v1}}^{*(\mathbf{h})} = \begin{bmatrix} -0.03 & 0 & 0 \\ 0 & 0.09 & 0 \\ 0 & 0 & 0 \end{bmatrix}.$$

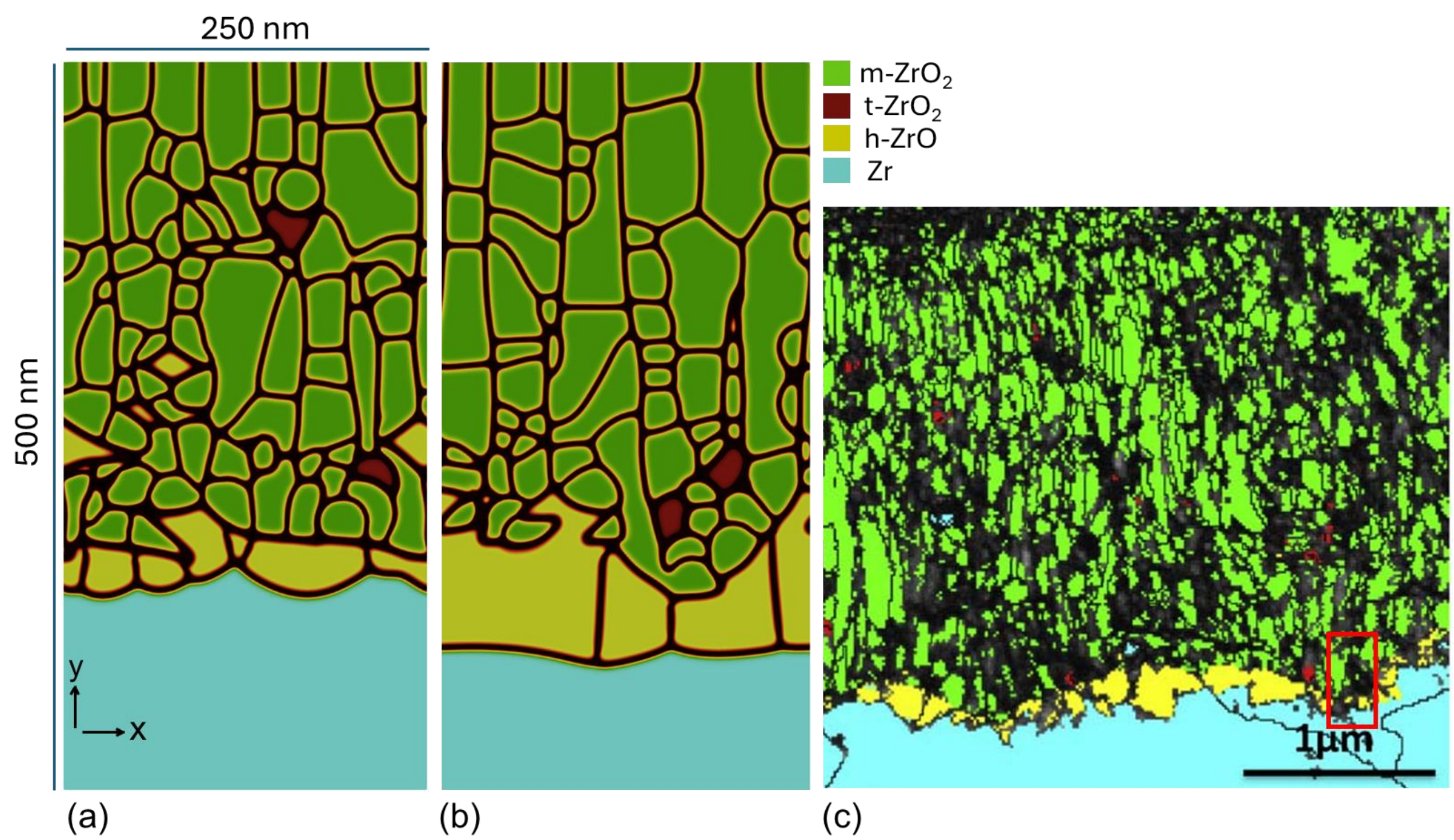

Figure 6-1. Comparison of simulated and experimental oxide microstructures. (a) Phase-field simulation employing the alternative orientation relationship between m-$ZrO_2$ and h-ZrO. (b) Phase-field simulation used in the main text. (c) Experimental micrograph.

The revised orientation relationship simulation (Figure 6-1 a) yields a t-$ZrO_2$ volume fraction comparable to both the main paper simulation and experiment. However, the h-ZrO grains are notably smaller than those in Figure 6-1b. The modified coherency strain increases elastic stabilization of h-ZrO within the surrounding monoclinic matrix, despite h-ZrO being chemically less favorable than the m-$ZrO_2$. This enhanced elastic stabilization promotes nucleation of small, equiaxed m-$ZrO_2$ grains surrounding embedded h-ZrO regions—a feature absent in the main paper simulation. The resulting microstructure exhibits a more columnar morphology with reduced stacking compared to Figure 6-1b, which we attribute to the more uniform spatial distribution of h-ZrO and t-$ZrO_2$ phases throughout the oxide layer which may impact the elastic forces that we believe led to the stacking (b).

### Sensitivity study of nucleation rate of m-$ZrO_2$ at h-ZrO/m-$ZrO_2$

To assess model robustness of the model in the main paper, we performed a sensitivity analysis examining the response of nucleation kinetics to ±10% variations in surface energy, diffusivity, and lattice mismatch (Figure 6-2 ). The effective nucleation rate shows minimal sensitivity to surface energy and eigenstrain variations, while diffusivity produces linear scaling. These results confirm that qualitative predictions regarding oxide microstructure evolution are insensitive to the parameters here within physically reasonable bounds.

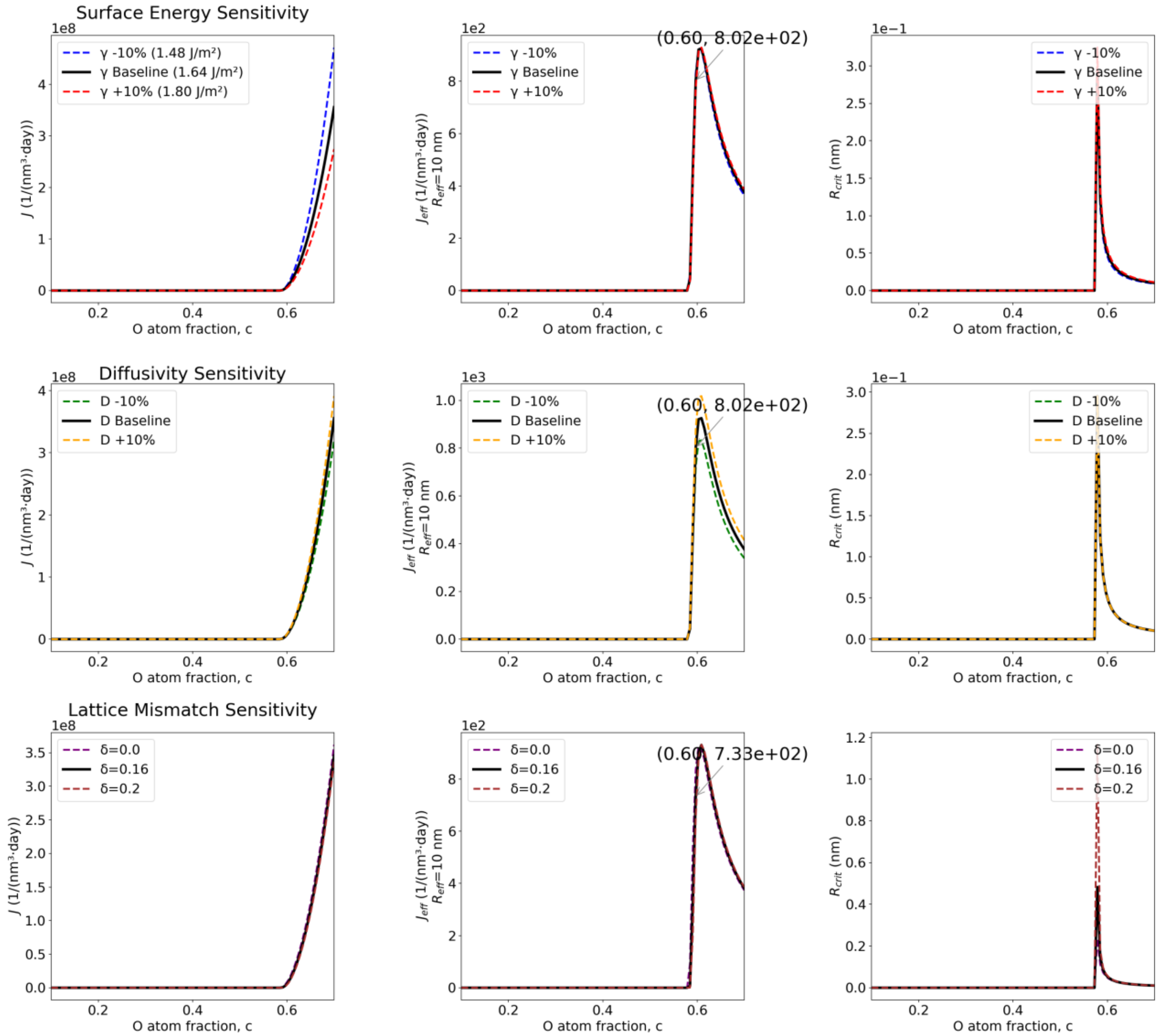

Figure 6-2. Evaluation of m-$ZrO_2$ nucleation rate sensitivity at the h-ZrO/m-$ZrO_2$ interface to ±10% perturbations in model parameters. Each row examines a different parameter while columns present (left) classical nucleation rate *J*, (center) effective nucleation rate *J*_eff for 10 nm nuclei, and (right) critical radius *R*_crit. (Top row) Surface energy: perturbations in γ_H-M cause significant shifts in *J* yet produce only minor changes in *J*_eff, as variations in the nucleation barrier and critical radius partially compensate. (Middle row) Diffusivity: *J* and *J*_eff respond proportionally to changes in diffusion coefficient, as expected from the linear role of diffusivity in attachment kinetics. (Bottom row) Lattice mismatch: the eigenstrain parameter δ spanning 0 to 0.2 has negligible influence on nucleation rates, confirming that the chemical free energy difference between h-ZrO and m-$ZrO_2$ overwhelms elastic strain energy contributions.

### Sensitivity study of nucleation rate of h-ZrO at Zr/m-$ZrO_2$

A similar sensitivity analysis was performed for m-$ZrO_2$ nucleation at the h-ZrO/m-$ZrO_2$ interface (Figure 5-3). The results mirror those at the metal-oxide interface: The effective nucleation rate exhibits minimal sensitivity to ±10% variations in surface energy despite substantial changes in the classical nucleation rate *J*, diffusivity produces proportional scaling, and lattice mismatch contributions remain negligible.

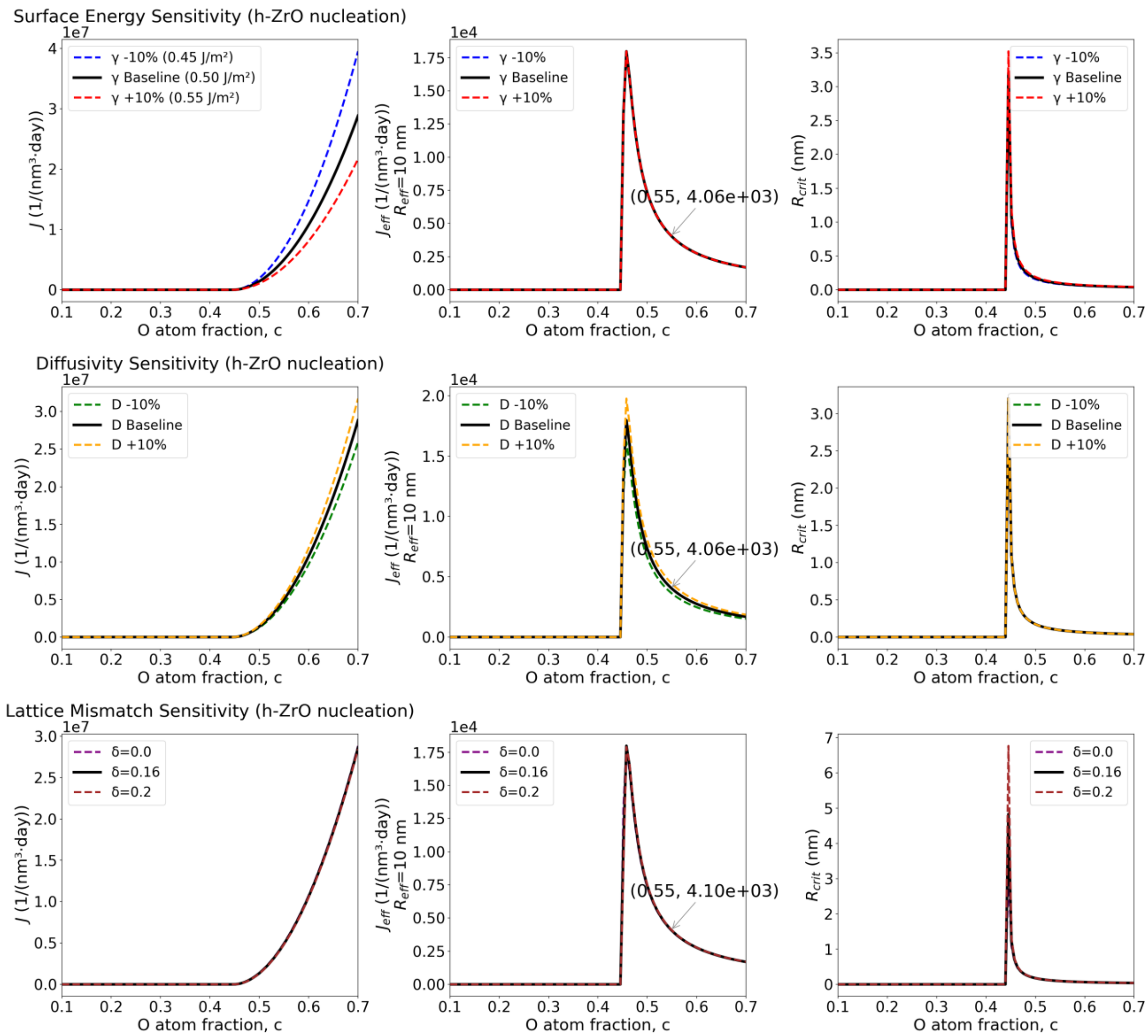

Figure 6-3. Sensitivity analysis of m-$ZrO_2$ nucleation kinetics at the h-ZrO/m-$ZrO_2$ interface to ±10% variations in key model parameters. Columns show (left) classical nucleation rate *J*, (center) effective nucleation rate *J*_eff for nuclei reaching 10 nm radius, and (right) critical nucleus radius *R*_crit. (Top row) Surface energy sensitivity: ±10% variation in γ_H-M produces substantial changes in *J* but minimal variation in *J*_eff due to compensating effects on *R*_crit. (Middle row) Diffusivity sensitivity: *J* and *J*_eff scale linearly with diffusion coefficient. (Bottom row) Lattice mismatch sensitivity: variation of eigenstrain parameter δ from 0 to 0.2 shows negligible effect on nucleation kinetics,

confirming that chemical driving forces dominate elastic contributions.

### Sensitivity study of nucleation rate of m-$ZrO_2$ at Zr/m-$ZrO_2$

The same analysis applied to m-$ZrO_2$ nucleation at the α-Zr/m-$ZrO_2$ interface yields consistent findings (Figure 6-3), confirming that the robustness of the effective nucleation rate to parameter variations holds across all nucleation sites considered in this work.

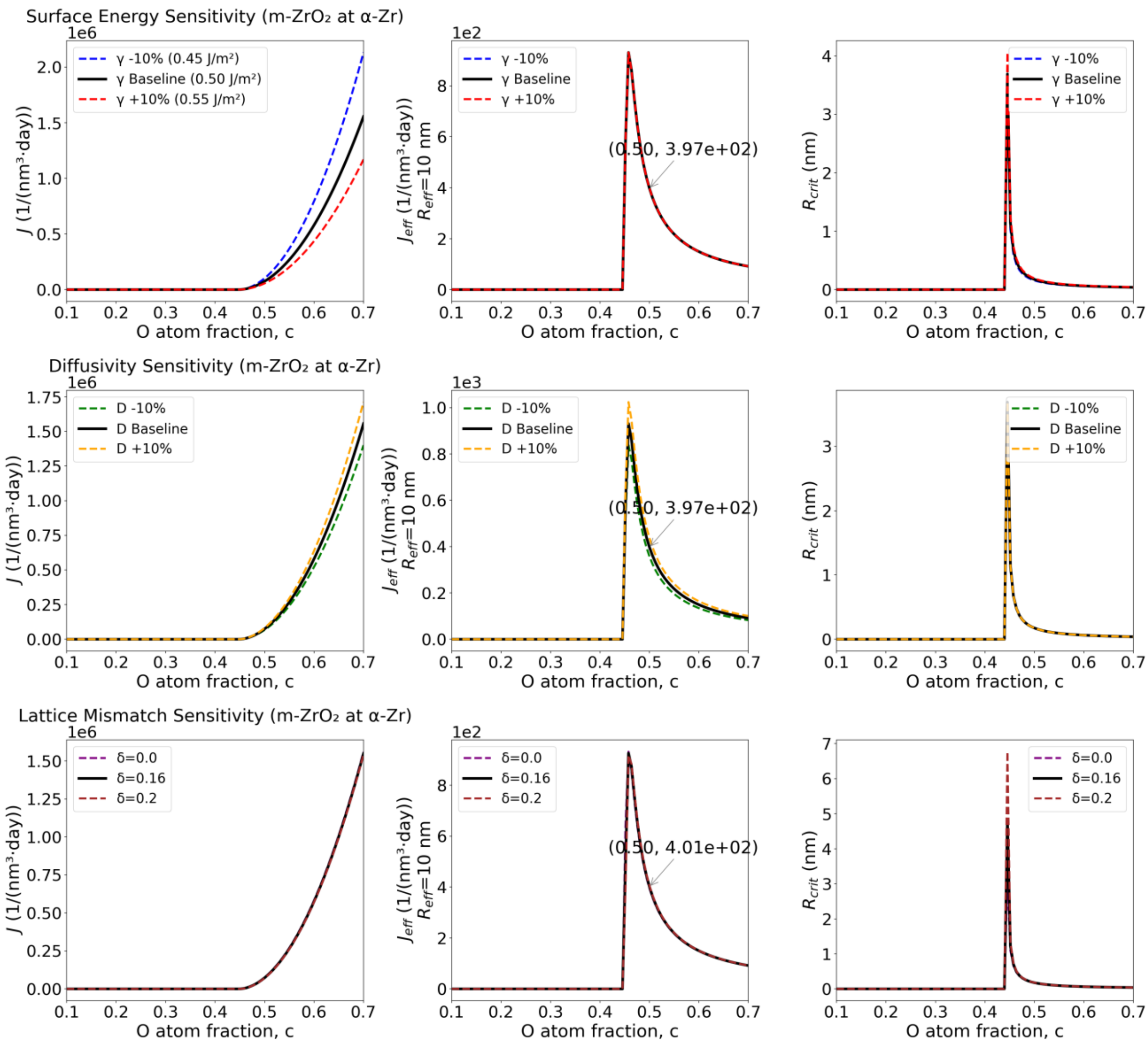

Figure 6-4. Parameter sensitivity of m-$ZrO_2$ nucleation kinetics at the α-Zr/m-$ZrO_2$ interface under ±10% variations in input parameters. Columns display (left) classical nucleation rate *J*, (center) effective nucleation rate *J*_eff corresponding to nuclei growing to 10 nm, and (right) critical nucleus radius *R*_crit. (Top row) Interfacial energy sensitivity:

±10% changes in γ_Zr-M yield large variations in *J* while *J*_eff remains nearly unchanged, reflecting the offsetting influence of critical radius on effective nucleation probability. (Middle row) Diffusion coefficient sensitivity: both *J* and *J*_eff exhibit proportional scaling with diffusivity, consistent with the linear dependence of atomic attachment kinetics. (Bottom row) Eigenstrain sensitivity: varying the lattice mismatch parameter δ from 0 to 0.2 produces negligible changes in nucleation behavior, indicating that elastic strain energy contributions are subordinate to chemical driving forces at this interface.